\documentclass[aip,cha]{revtex4-2}
\usepackage{graphicx}
\usepackage{dcolumn}
\usepackage{amsmath}
\usepackage{amssymb}
\usepackage{color}

\begin{document}

\title{Exploring the role of diffusive coupling in spatiotemporal chaos}
\author{A. Raj}
\affiliation{Department of Mechanical Engineering, Virginia Tech, Blacksburg, Virginia 24061}
\author{M. R. Paul}
\email{Corresponding author: mrp@vt.edu}
\affiliation{Department of Mechanical Engineering, Virginia Tech, Blacksburg, Virginia 24061}

\date{\today}

\begin{abstract}
We explore the chaotic dynamics of a large one-dimensional lattice of coupled maps with diffusive coupling of varying strength  using the covariant Lyapunov vectors (CLVs). Using a lattice of diffusively coupled quadratic maps we quantify the growth of spatial structures in the chaotic dynamics as the strength of diffusion is increased. When the diffusion strength is increased from zero, we find that the leading Lyapunov exponent decreases rapidly from a positive value to zero to yield a small window of periodic dynamics which is then followed by chaotic dynamics. For values of the diffusion strength beyond the window of periodic dynamics, the leading Lyapunov exponent does not vary significantly with the strength of diffusion with the exception of a small variation for the largest diffusion strengths we explore. The Lyapunov spectrum and fractal dimension are described analytically as a function of the diffusion strength using the eigenvalues of the coupling operator.  The spatial features of the CLVs are quantified and compared with the eigenvectors of the coupling operator.  The chaotic dynamics are composed entirely of physical modes for all of the conditions we explore. The leading CLV is highly localized and the localization decreases with increasing strength of the spatial coupling. The violation of the dominance of Oseledets splitting indicates that the entanglement of pairs of CLVs becomes more significant between neighboring CLVs as the strength of the diffusion is increased.
\end{abstract}

\maketitle

\begin{quotation}

Many important problems can be described as spatially extended systems with complex variations in both space and time. Examples include the dynamics of the weather and oceans and the turbulent fluid flow that occurs in a pipe. When spatially extended systems are driven strongly a common feature that emerges is spatiotemporal chaos where the temporal and spatial variations are aperiodic. In many cases of interest, the disorder in the system is generated locally which is then distributed spatially by coupling mechanisms  such as diffusion. We explore fundamental questions of spatiotemporal chaos using a nonlinear map as a source of local disorder. We consider a large number of identical nonlinear maps that are placed on a one-dimensional lattice where each map interacts with its nearest-neighbors  diffusively with a coupling strength that can be varied. We study the emergence of spatial order with increased coupling and quantify the spatiotemporal dynamics by computing the growth or decay of small perturbations to the dynamics. These perturbations are described using the covariant Lyapunov vectors which quantify the exponential growth or decay of the perturbations as well as their orientation in the tangent space. We find that the strength of diffusion has a significant effect upon the dynamics and upon the covariant Lyapunov vectors.  We use the covariant Lyapunov vectors to  provide new physical insights into the influence of spatial couplings on complex dynamics in space and time.

\end{quotation}

\section{Introduction}
\label{section:introduction}

Many large spatially-extended systems with spatial coupling among the degrees of freedom, that are driven far-from-equilibrium, exhibit rich dynamics that are often spatiotemporally chaotic~\cite{cross:1993}.  Examples include the dynamics of the atmosphere and oceans~\cite{chandrasekhar:1961}, fluid turbulence~\cite{pope:2000,kerswell:2005}, the dynamics of growing colonies of microorganisms~\cite{hagan:1981}, the complex spread of epidemics in highly mobile populations~\cite{hethcote:2000}, the intricate dynamics of a reacting species in a flow field~\cite{dewit:2020,mukherjee:2022}, and the complex patterns that emerge when a shallow layer of granular material is shaken~\cite{melo:1995,aranson:2006}.

In the theoretical description of many systems, it is fruitful to consider the chaotic dynamics to be generated locally which is then connected to different regions through mechanisms of spatial coupling. In many transport problems, important forms of spatial coupling are due to diffusive and convective phenomena.  For example, the diffusion of momentum in fluid dynamics and of mass in an advection-reaction-diffusion system. In complex networks it has been shown that long-range spatial connections can have a significant impact on the dynamics~\cite{watts:1998}. Electric and magnetic phenomena often yield complex and subtle spatial couplings in magnetohydrodynamical systems such as solar convection~\cite{ossendrijver:2003}. In the laboratory, the electroconvection of nematic liquid crystals yields striking chaotic dynamics and patterns due to the interplay of the electrical coupling with the fluid dynamics~\cite{krekhov:2015}. 

Exciting progress has been made in building our physical understanding of high-dimensional chaotic dynamics using a dynamical systems approach (\emph{cf.}~\cite{hof:2004,budanur:2017,suri:2017}). In this description, the dynamics is represented as a trajectory through a high (and possibly infinite) dimensional state space. This has led to powerful ideas such as exact coherent structures~\cite{waleffe:2001} which have provided insights into the skeleton or geometry of turbulence in state-space~\cite{hof:2004,suri:2017,cvitanovic:2013} and to the covariant Lyapunov vectors~\cite{ruelle:1979} (CLVs) which provide a rigorous description of the growth or decay of perturbations in the tangent space~\cite{ruelle:1979,kuptsov:2012,pikovsky:2016}. The CLVs have been used to determine the degree of hyperbolicity of chaotic dynamics~\cite{takeuchi:2011,inubushi:2012}, to demonstrate a decomposition of the tangent space into physical and transient modes~\cite{yang:2009,takeuchi:2011}, and to quantitatively describe the spatiotemporal dynamics of small perturbations to the nonlinear dynamics of fluid convection~\cite{xu:2016,xu:2018}.

However, from a dynamical systems perspective, partial differential equations are difficult to use for broad and fundamental studies because they formally represent an infinite dimensional state space~\cite{hopf:1948}. When partial differential equations are represented numerically using a computational approach, the dimension is no longer infinite but remains very large. For example, in a fundamental study tailored to minimally capture experimental conditions for a convecting fluid system~\cite{xu:2018} this dimension can be $\mathcal{O}(10^5)$. However, for more realistic applications, such as a model of the weather or the combustion of mixed gasses in an engine, the dimension would be much larger making such an approach prohibitive.

We conduct a fundamental study using the CLVs to investigate the spatiotemporal chaos of a lattice of coupled maps where we can tailor the local nonlinearity and the specific forms of the spatial couplings that are included. Coupled map lattices (CMLs) have a rich literature~\cite{kaneko:1989-stc,kaneko:1989-pattern,kaneko:1992} and have been used to gain new physical insights into a wide range of fundamental questions. This includes the dynamics of clouds~\cite{yanagita:1997}, models of fluid convection~\cite{yanagita:1995}, studies of nonequilibrium statistical mechanics~\cite{miller:1993,ohern:1996,egolf:2000:science}, the role of a conservation law on chaotic dynamics~\cite{bourzutschky:1992,grigoriev:1997:chaos,barbish:2023}, and the presence of hydrodynamic~\cite{yang:2013} or collective~\cite{takeuchi:2013} Lyapunov modes to name a few.

Our intention is not to quantitatively describe a specific physical system, or application, but to carefully explore the role of a local nonlinearity in the presence of diffusive coupling for a high-dimensional system exhibiting spatiotemporal chaos over a broad range of conditions.  The computational accessibility of a one-dimensional lattice of maps is a key element of our study that allows us to conduct such a broad and fundamental investigation. Using a one-dimensional lattice of coupled maps we have explored high-dimensional chaotic dynamics for a range of parameters and for very long times while also computing the entire spectrum of covariant Lyapunov vectors. The insights from our study can be used to guide future work on more complex systems such as the coupled nonlinear partial differential equations at the core of many important transport phenomena.

The outline of the remainder of the paper is the following. In \S\ref{section:approach} we describe our approach for using CMLs to study spatiotemporal chaos with diffusive coupling motivated by equations typically used to model transport phenomena. We then describe how the CLVs are computed for the specific case of interest. In \S{\ref{section:results}} we investigate the chaotic dynamics of a lattice for a wide range of diffusive coupling strengths. Lastly, in \S\ref{section:conclusion} we describe our conclusions.

\section{Approach}
\label{section:approach}

Our approach is to use a mapping of the form $u^{(n+1)} = f(u^{(n)})$ as a local source of chaotic dynamics where $f(u^{(n)})$ is a nonlinear mapping function and $u^{(n)}$ is a real variable specifying the state at discrete time $n$. Nonlinear maps have a rich literature~\cite{may:1976} and they are particularly attractive for their computational accessibility. Specifically, we use a centered logistic map, or quadratic map, given by 
\begin{equation}
u^{(n+1)} = r \left[ \frac{1}{4} - \left(u^{(n)}\right)^2 \right]
\label{eq:quadratic-map}
\end{equation}
where the real constant $r$ is the control parameter. The quadratic map exhibits the period-doubling route to chaos~\cite{may:1976} and, in our study, we use $r=2.8$ which yields a positive Lyapunov exponent for a single isolated map.

\begin{figure}[h!]
\vspace{1cm}
\begin{center}
\includegraphics[width=2.25in]{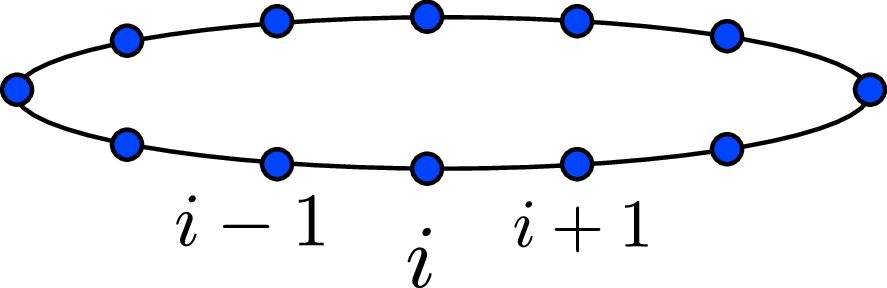}
\end{center}
\caption{A schematic of the one-dimensional periodic lattice where $i$ is the index of the lattice. Two nearest neighbors are indicated by $i\!+\!1$ and $i\!-\!1$, for the lattice shown there are $N\!=\!12$ sites for clarity of presentation, in our simulations we use a large lattice with $N\!=\!256$.}
\label{fig:geom}
\end{figure}

We use a one-dimensional spatially periodic lattice containing $N$ maps as shown schematically in Fig.~\ref{fig:geom}(a).  In our study we are interested in high-dimensional chaotic dynamics and use $N \!=\! 256$. Using our conventions, $i \!=\! 1, 2, \ldots, N$, where $i$ is the index representing the map location on the lattice. $u_i^{(n)}$ is the state of the map located at lattice $i$ at discrete time $n$.  The spatial periodicity of the lattice yields $u_i^{(n)} = u_{i+N}^{(n)}$ for all $i$ and $n$.

\subsection{Diffusive Coupling}

The diffusive coupling of the maps, located on the lattice, is added using a partial differential equation representation as a guide~\cite{kaneko:1993}. We start from the one-dimensional transport equation 
\begin{equation}
\frac{\partial u}{\partial t} = D \frac{\partial^2 u}{\partial x^2}
\label{eq:diffusion-pde}
\end{equation}
where $D$ is a diffusion coefficient, $x$ is space, and $t$ is time. Discretizing $u(x,t)$ in Eq.~(\ref{eq:diffusion-pde}) using a forward Euler time difference and a center difference for the diffusion term yields
\begin{equation}
\frac{u_i^{(n+1)} - u_i^{(n)}}{\Delta t} = D \left( \frac{u_{i+1}^{(n)} - 2 u_i^{(n)} + u_{i-1}^{(n)}}{\Delta x^2} \right)
\end{equation}
where $i$ represents the spatial discretization and $n$ the temporal discretization. Looking ahead to a lattice representation, we set $\Delta x \!=\! \Delta t \!=\! 1$, and rearrange to give 
\begin{equation}
u_i^{(n+1)} = u_i^{(n)} + D \left( u_{i+1}^{(n)} - 2 u_i^{(n)} + u_{i-1}^{(n)} \right).
\label{eq:diff1}
\end{equation}
We next include the generation of local disorder by imposing that each term on the right hand side of Eq.~(\ref{eq:diff1}) is first acted upon by the nonlinear mapping function to yield  
\begin{equation}
u_i^{(n+1)} = f(u_i^{(n)}) + D \left( f(u_{i+1}^{(n)}) - 2 f(u_i^{(n)}) + f(u_{i-1}^{(n)}) \right).
\label{eq:diff2}
\end{equation}
Equation~(\ref{eq:diff2}) indicates that the values of $u_i^{(n)}$ on the right hand side of Eq.~(\ref{eq:diff1}) are replaced with their individual values at the next time step $(n+1)$ after being acted upon by the mapping function which is independent of the dynamics of the neighbors. The diffusive coupling is then applied to these updated values. It is typical to use the convention $\epsilon = 2 D$ and to express this as 
\begin{equation}
u_i^{(n+1)} = (1 - \epsilon)f(u_i^{(n)}) + \frac{\epsilon}{2} \left( f(u_{i+1}^{(n)}) + f(u_{i-1}^{(n)}) \right)
\label{eq:diff3}
\end{equation}
where $\epsilon$ is the coupling constant representing the strength of the diffusion. Equation~(\ref{eq:diff3}) describes a lattice of nonlinear maps that are diffusively coupled using nearest neighbor coupling.

\subsection{Covariant Lyapunov vectors}

In the following we briefly describe the CLVs~\cite{ruelle:1979,pikovsky:2016} and the approach~\cite{ginelli:2007} used for their computation.  We provide only the essential details to describe our computations while focusing on the influence of the diffusive spatial coupling. CLVs emerge from a rigorous dynamical systems description~\cite{kuptsov:2012,ginelli:2013}, more details regarding their application to coupled map lattices using our conventions can be found in Ref.~\cite{barbish:2023}.

It will be useful to represent the lattice dynamics given by Eq.~(\ref{eq:diff3}) in a vector form where 
\begin{equation}
\vec{u}^{(n+1)} = \vec{g}(\vec{u}^{(n)}).
\label{eq:lattice-dynamics}
\end{equation}
In this notation $\vec{u}^{(n)}$ is an $N$ dimensional vector where each component is the value of the map at that lattice site at time step $n$. The components of $\vec{g}(\vec{u}^{(n)})$  are the right hand side of the dynamical equation evaluated at that lattice site.  $\vec{g}(\vec{u}^{(n)})$ contains the spatial coupling contributions and depends upon the values of the nearest neighbor lattice sites.  For diffusive spatial coupling $\vec{g}(\vec{u}^{(n)})$ is a linear function of $\vec{f} (\vec{u}^{(n)})$.

The dynamics of the $k$th small perturbation, $\delta \vec{u}_k$, to the lattice values, $\vec{u}$, are described by the tangent space equations
\begin{equation}
\delta \vec{u}_k^{(n+1)} = \mathbf J^{(n)} \delta \vec{u}_k^{(n)}
\label{eq:pert-dynamics}
\end{equation}
where $k=1,2, \ldots,N_\lambda$ is the index for the perturbation and $N_\lambda$ is the total number of perturbations being considered. The number of perturbations determines the number of CLVs that can be calculated. For CMLs, the maximum number of CLVs that can be calculated is equal to the dimension of the system $N$. We will refer to $k$ as the Lyapunov index and in our study we have used $N_\lambda = N$.   $\mathbf{J}^{(n)}$ is the $N \times N$ Jacobian matrix defined by
\begin{equation}
\mathbf{J}^{(n)} = \left( \frac{\partial \vec{g}(\vec{u})}{\partial \vec{u}} \right)^{(n)}.
\end{equation}
The $k$th linear equation describing the perturbation dynamics can be expressed explicitly as
\begin{equation}
\delta u_{k,i}^{(n+1)} = \left(1-\epsilon \right) f'\left(u_{i}^{(n)} \right) \delta u_{k,i}^{(n)} + \frac{\epsilon}{2} f'\left(u_{i+1}^{(n)} \right) \delta u_{k,i+1}^{(n)} + \frac{\epsilon}{2} f'\left(u_{i-1}^{(n)} \right) \delta u_{k,i-1}^{(n)}
\label{eq:diff-pert}
\end{equation}
where $\delta u_{k,i}$ is the value of the $k$th perturbation at lattice site $i$ and $f'(u)$ is the derivative of $f(u)$ given by the right hand side of Eq.~(\ref{eq:quadratic-map}). Therefore, the CLV calculation requires the iteration of Eq.~(\ref{eq:diff3}) with $N_\lambda$ instances of Eq.~(\ref{eq:diff-pert}).

The essential ideas for computing the CLVs are the following. First, a lattice with small random initial conditions is iterated forward in time using  Eq.~(\ref{eq:lattice-dynamics}). The dynamics are iterated for a long time, $n \sim \mathcal{O}(10^6)$, to allow for all initial transients to decay. At this time, the lattice dynamics are continued while also iterating $N_\lambda$ instances of Eq.~(\ref{eq:diff-pert}).

The perturbation vectors $\delta \vec{u}_k^{(n)}$ can be arranged as the columns of a $N_\lambda \times N_\lambda$ matrix. Periodically in time, a $\mathbf{QR}$ decomposition~\cite{trefethen:1997} of this matrix is computed to yield the $N_\lambda \times N_\lambda$ matrices $\mathbf{Q}$ and $\mathbf{R}$. The columns of $\mathbf{Q}$ are the orthonormal Gram-Schmidt vectors $\vec{q}_k^{\,(n)}$ and the upper triangular matrix $\mathbf{R}$ contains the expansion and contraction coefficients on the diagonal and the remaining off-diagonal elements contain information that will be useful in determining the directions of the CLVs. The natural logarithm of the diagonal elements of $\mathbf{R}$ yield values of the instantaneous Gram-Schmidt Lyapunov exponents.

This forward time evolution is continued for a sufficiently long time while computing and storing the periodically computed $\mathbf{Q}$ and $\mathbf{R}$ matrices.   At this point, the equations are evolved backwards in time while the CLVs, $\vec{v}_k^{(n)}$, are constructed from a linear combination of the previously computed Gram-Schmidt vectors, $\vec{q}_k^{\,(n)}$, contained in the columns of the stored $\mathbf{Q}$ matrices which requires use of the stored $\mathbf{R}$ matrices. The matrix of coefficients used to construct the CLVs from the $\vec{q}_k^{\,(n)}$ are computed during this backward time evolution using the combination matrix $\mathbf{C}$ which is defined as 
\begin{equation}
C_{j,i}^{(n-1)} = \left[ R_{j,i}^{(n)} \right]^{-1} C_{j,i}^{(n)}
\label{eq:cmatrix}
\end{equation}
where $\mathbf{R}^{-1}$ is the inverse of $\mathbf{R}$. We note that it is straightforward to solve Eq.~(\ref{eq:cmatrix}) for the combination matrix using a back substitution approach which does not require the explicit computation of the inverse of $\mathbf{R}$.

The CLVs are then computed from the combination matrix as 
\begin{equation}
\vec{v}_k^{(n)} = \sum_{j=1}^{k} C_{j,k}^{(n)} \vec{q}_j^{\,(n)}
\label{eq:clv}
\end{equation}
where each of the $N_\lambda$ CLVs is a normalized vector with $N$ components. We emphasize that the CLVs are not orthogonal with respect to each other and their direction in the tangent space is physically relevant. The leading CLV, $\vec{v}_1$, and the leading Gram-Schmidt vector, $\vec{q}_1$, are identical.  However, the remaining vectors for $k \!=\! 2, 3, \ldots, N_\lambda$ are different and it is the CLVs which should be used to probe the dynamics in the tangent space. It is important to also note that the finite time Gram-Schmidt Lyapunov exponents and the finite time covariant Lyapunov exponents (CLEs) are not the same (for $k \! \ge \! 2$), although they do agree in the infinite time limit. As a result, the CLEs should be used when computing diagnostics such as the violation of the domination of Oseledets splitting (DOS)~\cite{takeuchi:2011}.

Our general numerical approach is the following. We use the quadratic map given by Eq.~(\ref{eq:quadratic-map}) on a periodic lattice with $N\!=\!256$ lattice sites with the control parameter set to $r \!=\! 2.8$ which yields chaotic dynamics for a single isolated map. The Lyapunov exponent for a single isolated map $\lambda_0$ is given by 
\begin{equation}
\lambda_0 = \lim_{n \rightarrow \infty} \left( \frac{1}{n} \sum_{i=0}^{n-1} \ln \left| f'(u^{(i)}) \right| \right)
\end{equation}
and, for $r \! = \! 2.8$, this yields $\lambda_0 \!=\! 0.6034$.

Each simulation starts from small random initial conditions which is then iterated for $n \gtrsim 10^6$ time steps to allow for initial transients to decay. At this point we continue iterating the CML while also iterating $N_\lambda \!=\! 256$ perturbation equations. We then iterate the entire $N_\lambda + 1$ system of equations forward in time for another $n \gtrsim 10^4$ time steps while computing a $\mathbf{QR}$ decomposition every time step and storing the $\mathbf{Q}$ and $\mathbf{R}$ matrices.

We next iterate Eq.~(\ref{eq:cmatrix}) backwards in time for $n \!\gtrsim\! 4 \times 10^3$ to compute the combination matrix. Once the combination matrix has converged, the CLVs are assembled from the stored Gram-Schmidt vectors using Eq.~(\ref{eq:clv}). We typically compute the CLVs and CLEs for at least $10^3$  time steps which we then use in our analysis. The CLEs are computed by iterating forward in time each of the CLVs by a single time step and computing their growth or decay, this is repeated for all of the time steps for which the $N_\lambda$ CLVs have been calculated. We have conducted many tests using different periods of time for the different steps and have found that our results are not sensitive to these changes.

\section{Results}
\label{section:results}

\subsection{Spatiotemporal Dynamics of the Lattice}
\label{section:diffusion}

\begin{figure}[h!]
\begin{center}
\includegraphics[width=3.25in]{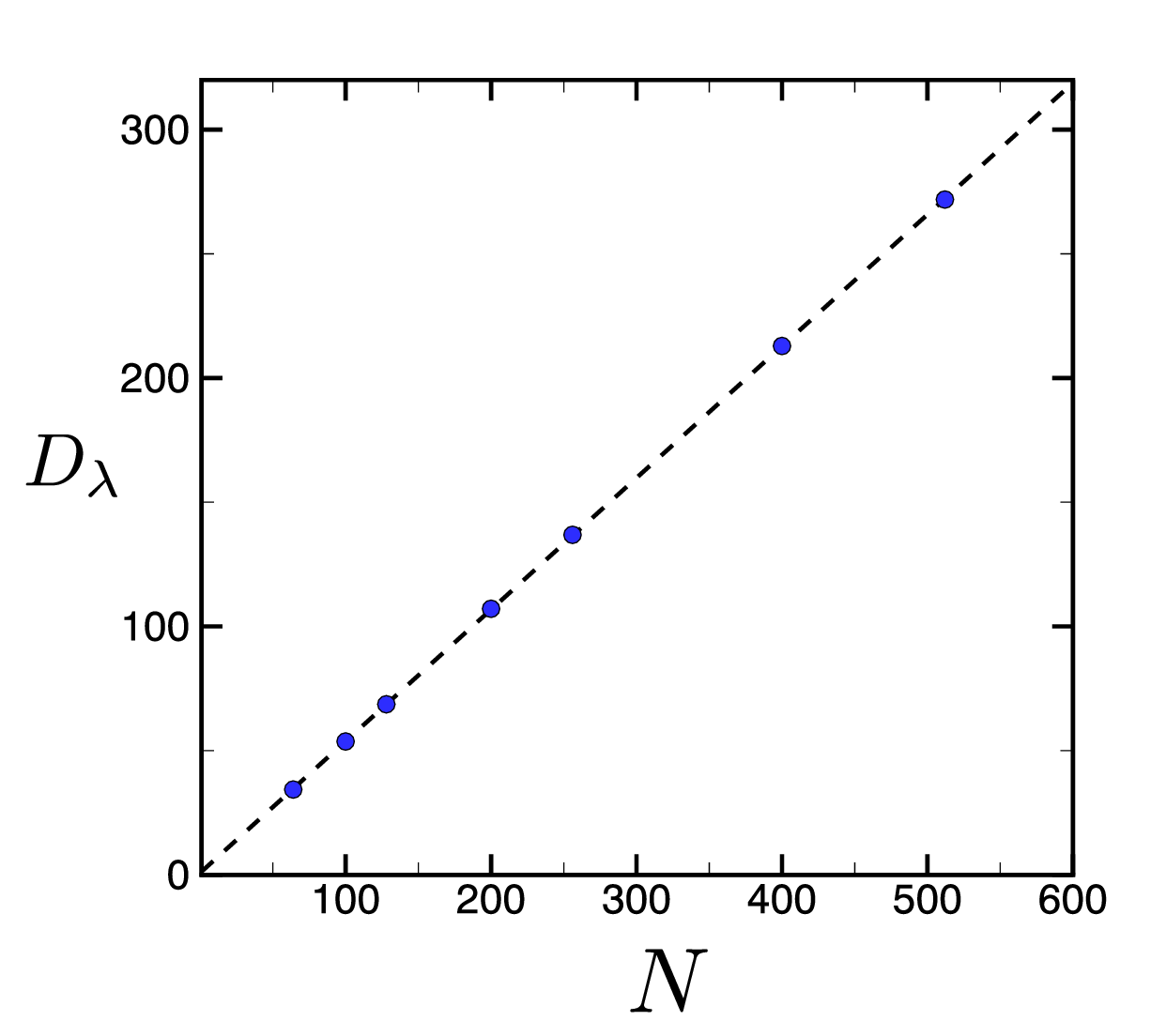}
\end{center}
\caption{The variation of the fractal dimension $D_\lambda$ with the lattice size $N$ for $\epsilon \!=\! 0.7$. The solid line is a linear curve fit through the results, $D_\lambda \!=\! 0.7777 \!+\! 0.5301 N$, where $D_\lambda \propto N$  indicates that the dynamics are extensively chaotic for $N \gtrsim 64$ with a dimension density of $\delta_\lambda \!\approx\! 1/2$.}
\label{fig:dlambda}
\end{figure}

We are interested in exploring high-dimensional chaotic dynamics for large spatially extended systems which requires a large value of $N$. In order to select a sufficiently large value of $N$, while taking care to not use a value of $N$ that is unnecessarily large,  we choose its value such that the dynamics are well within the extensively chaotic regime. Extensive chaos occurs when the fractal dimension (or Kaplan-Yorke dimension)\cite{kaplan:1979,farmer:1983} is proportional to the system size\cite{ruelle:1982}. In our case of a one-dimensional lattice of coupled maps, extensive chaos occurs when $D_\lambda \propto N$. As a representative example, in Fig.~\ref{fig:dlambda} we show the variation of $D_\lambda$ with $N$ for the case where $\epsilon \!=\! 0.7$ and $r \!=\! 2.8$. The dashed line is a linear curve fit through the data points where $D_\lambda \!=\! 0.7777 \!+\! 0.5301 N$ which clearly indicates extensive chaos for $N \gtrsim 64$.

These results also indicate that the dimension density $\delta_\lambda \!=\! D_\lambda/N$ of the dynamics is $\delta_\lambda \approx 1/2$ suggesting that the addition of approximately two lattice sites results is required to add a single chaotic degree of freedom to the dynamics while holding all other parameters constant. For $N \!<\! 64$ the variation of the dynamics with $N$ is a complicated mixture of periodic and chaotic dynamics and we have not explored this range further. With these results in mind we have chosen $N\!=\!256$ for all of our simulations in order for the chaotic dynamics to be in the extensive regime.

We first investigate the dynamics of a diffusively coupled lattice as a function of the diffusion strength $\epsilon$ over the range $0 \!\le\! \epsilon \!\le\! 1$. For $\epsilon\!=\!0$, this is simply a lattice of $N$ uncoupled maps which yields a Lyapunov exponent spectrum of $\lambda_k = \lambda_0$ for all $k$. As $\epsilon$ increases beyond zero, spatial structure emerges as the dynamics at the different lattice sites become coupled to yield rich spatiotemporal dynamics.

\begin{figure}[h!]
\begin{center}
\includegraphics[width=2.5in]{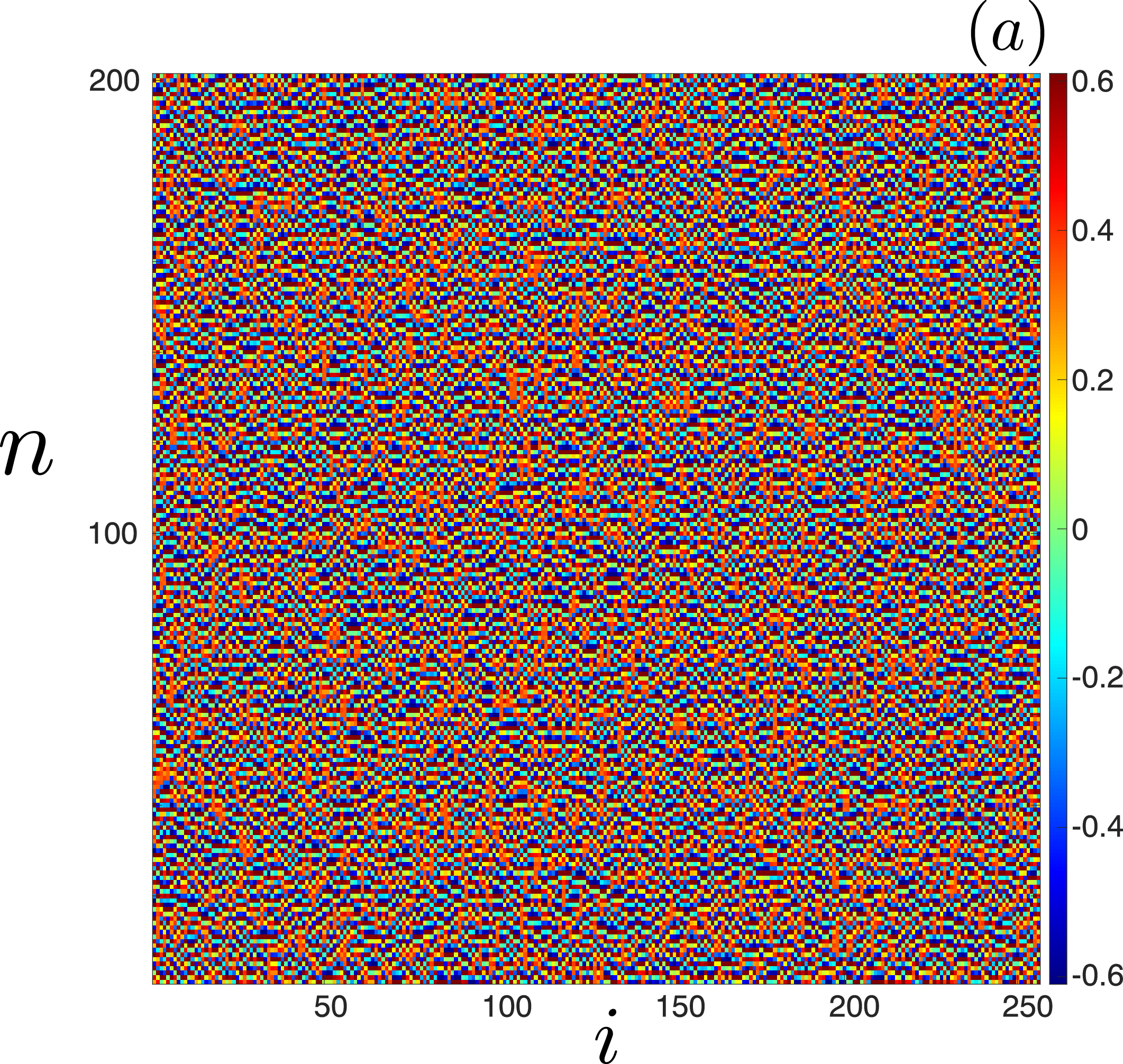}
\includegraphics[width=2.5in]{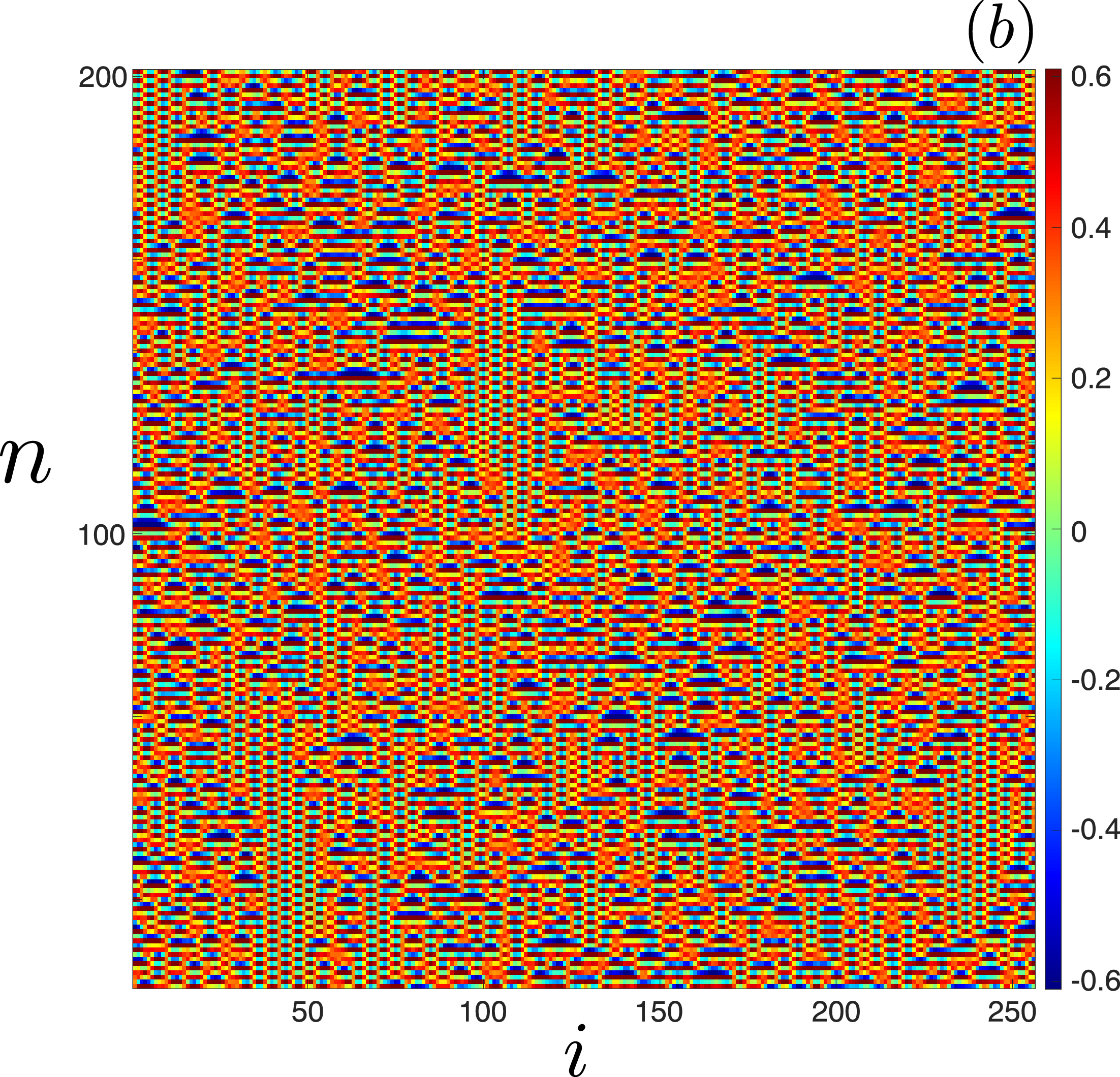}
\end{center}
\caption{Space-time plots of $u_i^{(n)}$. 200 time steps are shown after $10^6$ iterations beginning from random initial conditions. ($a$)~$\epsilon \!=\! 0.05$, ($b$)~$\epsilon \!=\! 0.7$.}
\label{fig:spacetime}
\end{figure}

Figure~\ref{fig:spacetime} shows space-time plots of the dynamics for a small value of the diffusion strength, $(a)$ $\epsilon\!=\!0.05$, and in~($b$) for a larger value $\epsilon\!=\!0.7$. The abscissa is the lattice index $i$, the ordinate is the time step $n$, and the color contours are of the state of the maps at each lattice site $u_i^{(n)}$. Each lattice began from the same random initial conditions and was iterated for $10^6$ time steps, the dynamics shown are the 200 time steps after this initial warm up.

Figure~\ref{fig:spacetime}($a$)-($b$) indicates complex spatiotemporal dynamics with larger spatial structures for the case of stronger diffusion.  As a measure of the spatial structure we have computed the two-point spatial correlation length $\xi_0$ for these dynamics. We estimate $\xi_0$ as the first zero crossing of the spatial correlation function. The value of the zero crossing is determined using a linear interpolation at the location of the first change of sign in the spatial correlation. $\xi_0$ is then estimated as this value rounded to the nearest integer in order to provide a result in the more meaningful lattice spacing units. Using this approach yields $\xi_0 \approx 2$ lattice spacings for $\epsilon \!=\! 0.05$ and $\xi_0 \!\approx\! 5$ lattice spacings for $\epsilon \!=\! 0.7$ which quantifies the growth of spatial structure with increasing diffusion strength. 

\begin{figure}[h!]
\begin{center}
\includegraphics[width=2.25in]{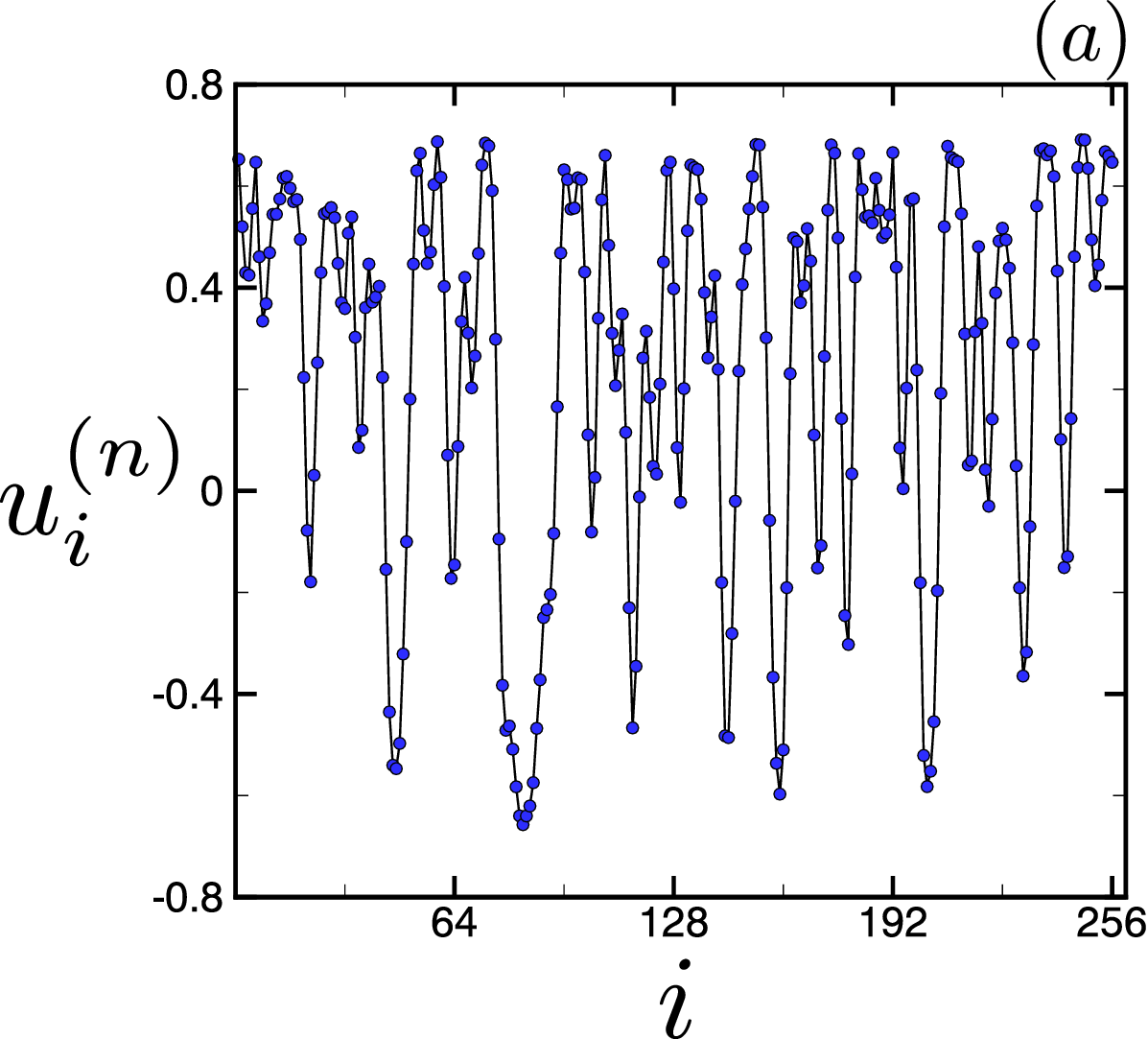}
\includegraphics[width=2.25in]{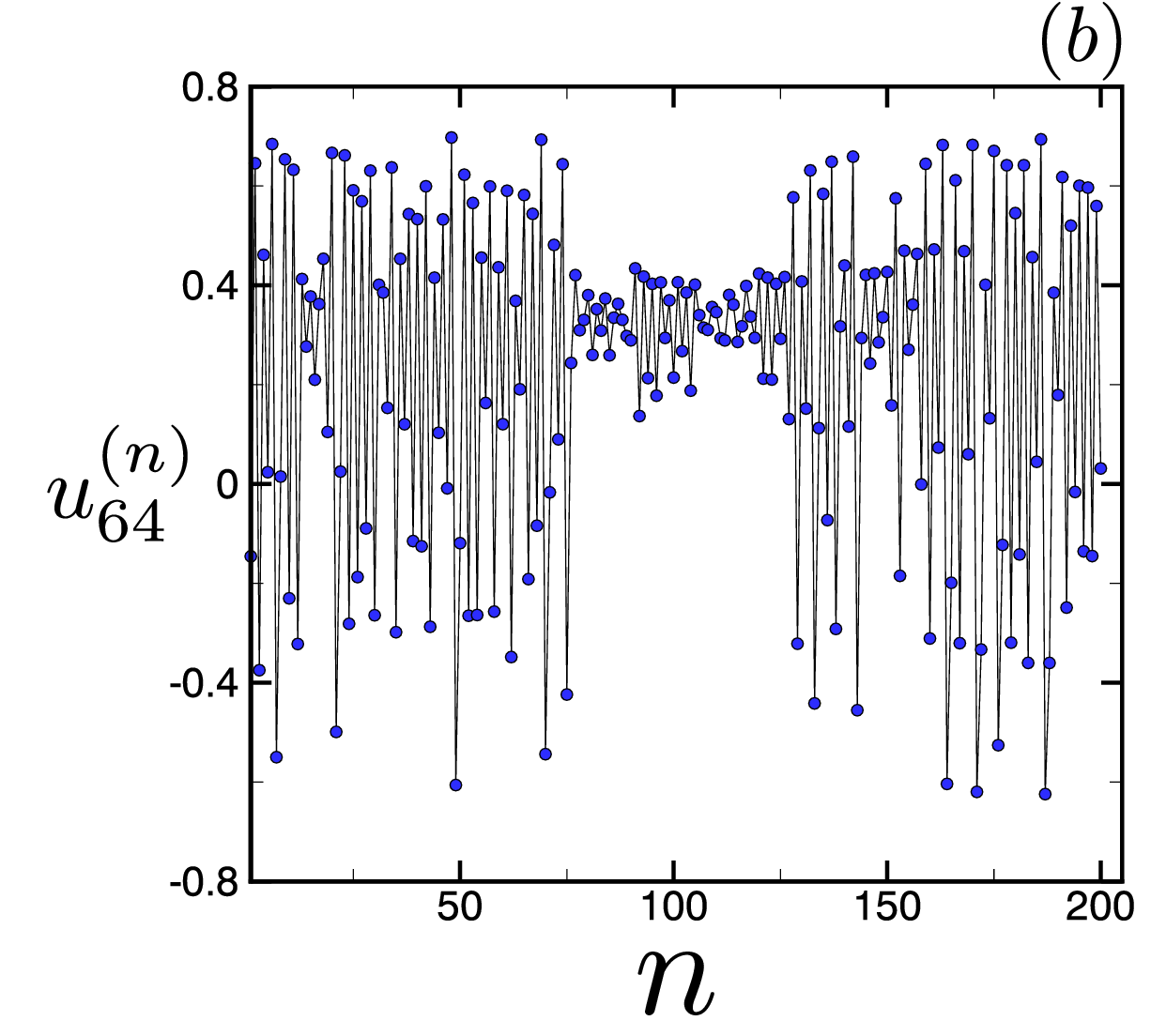}
\end{center}
\caption{The spatial structure and time variation of the lattice dynamics ($\epsilon \!=\! 0.7$). ($a$)~The state of the entire lattice at one instant of time where $n \!=\! 2 \times 10^6$. ($b$) The time variation of lattice site $i\!=\!64$ for 200 time steps. The sequence of time steps in~($b$) begins from the state shown in~($a$).}
\label{fig:lattice-dynamics}
\end{figure}

The dynamics for $\epsilon \!=\! 0.7$ are shown in additional detail in Fig.~\ref{fig:lattice-dynamics} to highlight the spatial structure in~($a$) and the temporal variation in~($b$). Figure~\ref{fig:lattice-dynamics}($a$) shows the values of maps of the entire lattice at a single instant of time. In this representation, the spatial structure of the dynamics aligns with the estimate of a correlation length of approximately 5 lattice spacings.  Figure~\ref{fig:lattice-dynamics}($b$) illustrates the variation in time of the state of a single map at a single lattice site, the site chosen is $i\!=\!64$, and the dynamics are shown for 200 time steps. In Fig.~\ref{fig:lattice-dynamics}($a$)-($b$) the solid lines are only included as a guide for viewing the dynamics. Figure~\ref{fig:lattice-dynamics} also indicates that the average of the dynamics over the lattice is in general positive, this is  reflected in Fig.~\ref{fig:spacetime} by the overall red-orange color of the space-time plots.

\subsection{Lyapunov spectra}
\label{section:lyapunov-spectra}

\begin{figure}[h!]
\begin{center}
\includegraphics[width=2.5in]{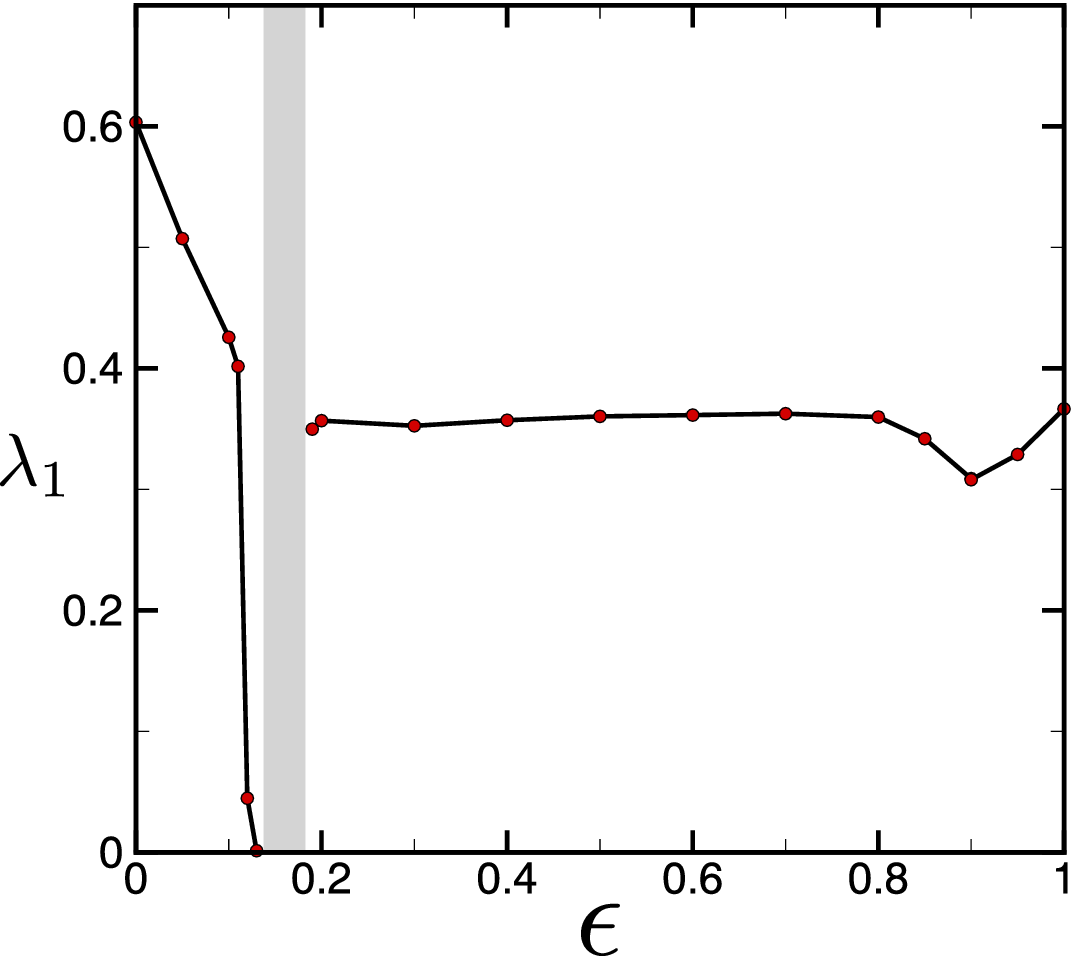}
\end{center}
\caption{The variation of the leading Lyapunov exponent $\lambda_1$ with diffusion strength $\epsilon$. The grey shaded region indicates a window of periodic dynamics.}
\label{fig:lyap1-with-epsilon}
\end{figure}

The leading Lyapunov exponent $\lambda_1$ is an important measure of the lattice dynamics and a positive value is the defining characteristic of chaos. The variation of $\lambda_1$ with $\epsilon$ is shown in Fig.~\ref{fig:lyap1-with-epsilon}. For small values of diffusion, $\epsilon \!\lesssim\! 0.13$, $\lambda_1$ decays rapidly towards zero with increasing $\epsilon$ to yield a window of periodic dynamics indicated by the grey shaded region. The window of periodicity is $0.14 \! \lesssim \! \epsilon \! \lesssim \! 0.18$.  For $\epsilon \!\gtrsim\! 0.19$ the dynamics are again chaotic for the entire range we have calculated. There is a large plateau for $0.19 \lesssim \!\epsilon\! \lesssim 0.8$ where the leading Lyapunov exponent remains nearly constant at $\lambda_1 \!\approx\! 0.35$.  For large diffusion strengths, $\epsilon \!\gtrsim\! 0.8$, there is a small dip in the value of $\lambda_1$ with the minimum occurring at $\epsilon \!=\! 0.9$.

Overall, Fig.~\ref{fig:lyap1-with-epsilon} indicates that a small amount of diffusion has a significant effect on the chaotic dynamics resulting in a decreasing value of the leading Lyapunov exponent. However, for larger values of $\epsilon$, after the window of periodic dynamics, increased diffusion has a negligible effect on $\lambda_1$. These four regions of: chaos at small values of diffusion; a window of periodicity; a region of chaos with a nearly constant $\lambda_1$; and a small region which contains a local minimum of $\lambda_1$ align with the four regions found by Shabunin~\cite{shabunin:2021} for a lattice of diffusively coupled logistic maps that were analyzed using the mutual coherence length.

We are interested in the chaotic dynamics that occur for the larger values of $\epsilon$ which is anticipated to be more representative of dissipative spatially extended systems. We have not probed further the transitions between periodic and chaotic dynamics that occur for the smaller values of $\epsilon$. Rather we focus on the larger diffusion cases where the spatial coupling is more significant.

\begin{figure}[h!]
\begin{center}
\includegraphics[width=2.25in]{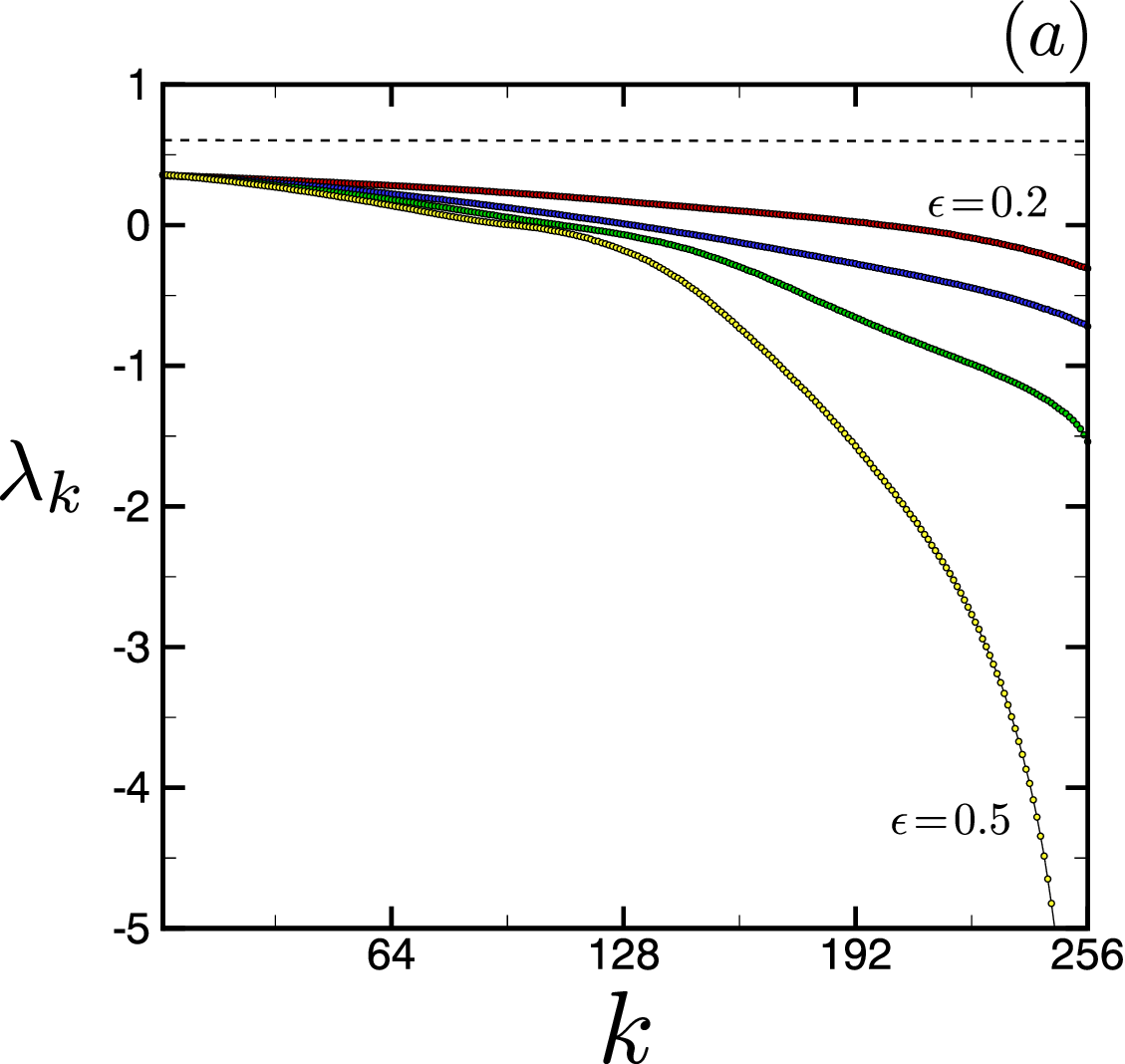}
\includegraphics[width=2.25in]{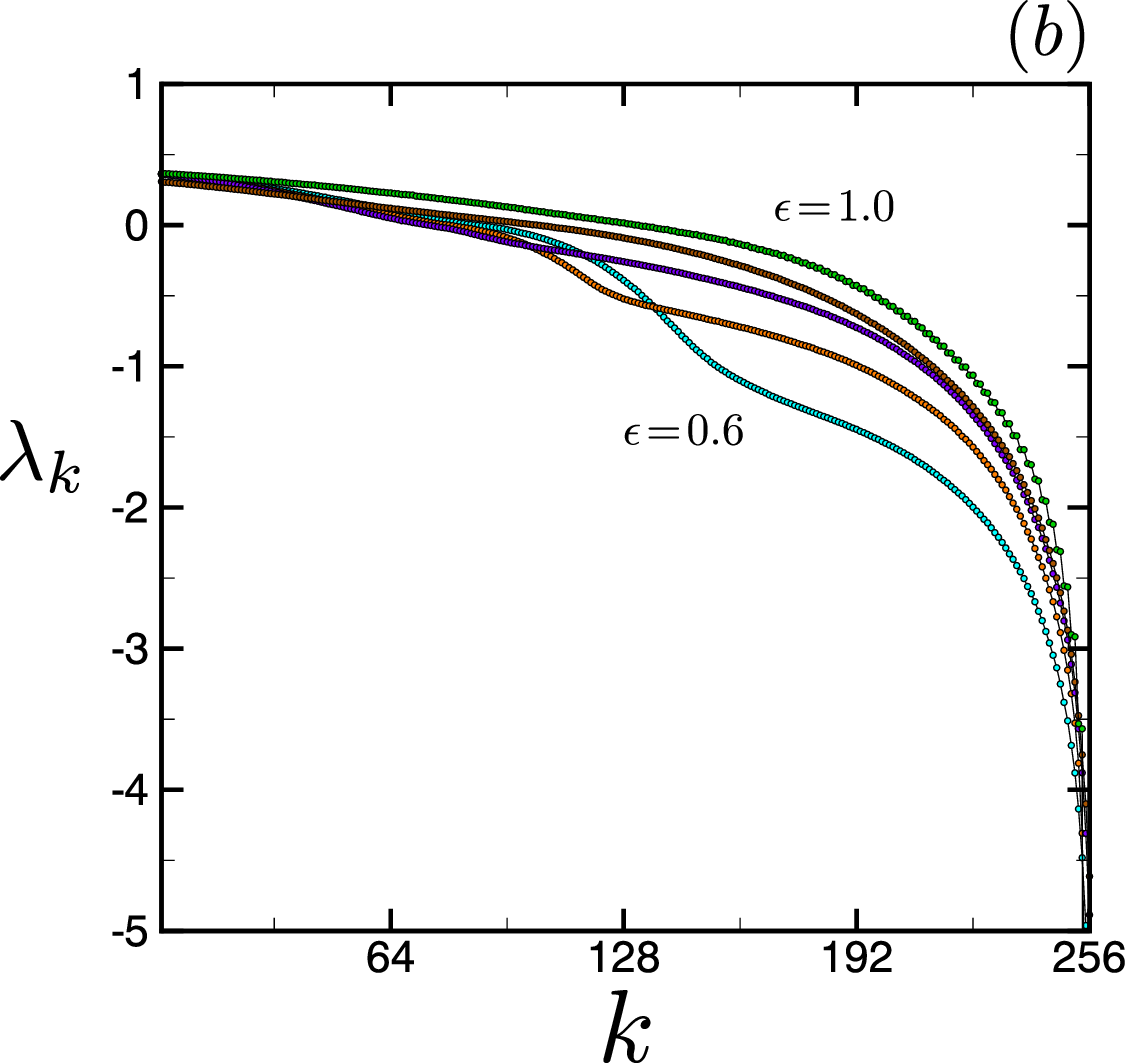}
\end{center}
\caption{The variation of the Lyapunov exponent spectrum $\lambda_k$ as a function of diffusion strength $\epsilon$. ($a$)~$\lambda_k$ decreases with increasing $\epsilon$ for $\epsilon \lesssim 0.5$.  The horizontal dashed line shows $\lambda_k=\lambda_0$ for the case of uncoupled maps ($\epsilon=0$).  ($b$)~For $\epsilon \gtrsim 0.5$, the region of large negative $\lambda_k$  increases with increasing values of $\epsilon$. Spectra are shown in increments of $\Delta \epsilon = 0.1$, unlabeled curves are in order between the bounding values.}
\label{fig:lyap-spec-diff-with-epsilon}
\end{figure}

The entire spectrum of the Lyapunov exponents is shown in Fig.~\ref{fig:lyap-spec-diff-with-epsilon} over the range of diffusion strengths $0.2 \!\le\! \epsilon \!\le\! 1.0$. The diffusion strength has a significant effect upon the $\lambda_k$ with a large Lyapunov index $k$. In Fig.~\ref{fig:lyap-spec-diff-with-epsilon}(a) the spectra are shown for $\epsilon \!\le\! 0.5$. The horizontal dashed line is the Lyapunov spectrum for the case where $\epsilon = 0$, corresponding to a lattice without spatial coupling, yielding $\lambda_k \!=\! \lambda_0$ for all $k$.

For small values of $k$, the $\lambda_k$ are not strongly affected by the increasing $\epsilon$ which is in agreement with the small variations of $\lambda_1$ with $\epsilon$ shown in Fig.~\ref{fig:lyap1-with-epsilon}.  For larger values of $k$, an increasing $\epsilon$ results in a significant reduction in $\lambda_k$.  The effect is most pronounced for the negative Lyapunov exponents, $\lambda_k \!<\! 0$, which become more negative with increasing $\epsilon$.

This general trend is in agreement with the expectation that diffusive coupling tends to damp out small scale dynamics. However, for larger values of $\epsilon$, the large positive $\lambda_k$ ($k \!\lesssim\! 10$) and the large negative $\lambda_k$ ($k \!\gtrsim\! 246$) remain relatively unaffected while the portion of the Lyapunov spectrum that lies between these two extremes \emph{increases} significantly.  This results in a lifting of the Lyapunov spectrum over this region.

Insight into these variations of $\lambda_k$ can be obtained by examining the eigenvalues of the diffusive coupling operator as discussed by Takeuchi \emph{et al.}~\cite{takeuchi:2011}. For the case of nearest neighbor diffusive coupling, the spatial coupling is linear. As a result, $\vec{g}(\vec{u}^{(n)})$ of Eq.~(\ref{eq:lattice-dynamics}) is a linear function of $\vec{f} (\vec{u}^{(n)})$ and the lattice dynamics can be expressed as
\begin{equation}
\vec{u}^{(n+1)} = \mathbf{A}_c \vec{f}({\vec{u}}^{(n)})
\end{equation}
where $\mathbf{A}_c$ is an $N \!\times\! N$ coupling matrix with constant coefficients representing the diffusive coupling that is independent of the particular nonlinear mapping function that is used. In our notation, $\vec{f}({\vec{u}}^{(n)})$ is the vector representation of the nonlinear mapping function $f$ applied at each lattice site and discrete time $n$. The coupling matrix has $N$ real eigenvalues $\Lambda_k$ for $k \!=\! 1, 2, \ldots, N$ which have been arranged such that $|\Lambda_k|$ is in descending order.  For simplicity of the discussion we will assume that $N$ is an even number although this is not required.   An estimate of the influence of the diffusive coupling on the Lyapunov exponents can be obtained~\cite{takeuchi:2011} as $\lambda_k \!=\! \lambda_0 \!+\! \ln{ | \Lambda_k|}$ for $k \!=\! 1, 2, \ldots, N_\lambda$. Using this description, excellent agreement was found when comparing with numerical results for $\lambda_k$ of diffusively coupled tent maps for a case of large diffusion strength~\cite{takeuchi:2011}.

In this description, the shape of the Lyapunov spectrum is determined entirely by the variation of the eigenvalues $\Lambda_k$ where the leading Lyapunov exponent is $\lambda_1 \!=\! \lambda_0$ since the leading eigenvalue is $\Lambda_1 \!=\! 1$. However, as shown in Fig.~\ref{fig:lyap1-with-epsilon}, for the range of parameters we are exploring, the value of $\lambda_1$ is reduced by the presence of diffusive coupling. In order to account for the effect that this reduction of $\lambda_1$ has on the Lyapunov spectrum we will estimate $\lambda_k$ as    
\begin{equation}
\lambda_k = \lambda_1 + \ln{ | \Lambda_k|}
\label{eq:lyap-theory}
\end{equation}
where $\lambda_1$ must be calculated numerically for the diffusively coupled lattice.

For nearest neighbor diffusive coupling with periodic boundary conditions, analytical expressions for the eigenvalues $\Lambda_k$ and eigenvectors $\vec{\xi}_k$ of the circulant coupling matrix $\mathbf{A}_c$ can be found using Fourier methods~\cite{takeuchi:2011}. The eigenvalues can be expressed as 
\begin{equation}
\Lambda_{k'} = 1 - \epsilon \left[ 1 - \cos \left( \frac{2 \pi (k'-1)}{N} \right) \right]
\label{eq:eigenvalues-unsorted}
\end{equation}
for $k'=1, 2, \ldots, N$. The index $k'$ in Eq.~(\ref{eq:eigenvalues-unsorted}) indicates that the eigenvalues, when represented this way, are not sorted by decreasing magnitude as reflected by the cosine dependence with respect to $k'$. It is clear from the symmetry of Eq.~(\ref{eq:eigenvalues-unsorted}) that the eigenvalues, when sorted, will occur in pairs except for the single values at the maximum ($k' \!=\!1$) and minimum ($k' \!=\! N/2 \!+\! 1$). It will be useful to first discuss the variation of the unsorted eigenvalues with respect to $\epsilon$.

\begin{figure}[h!]
\begin{center}
\includegraphics[width=2.25in]{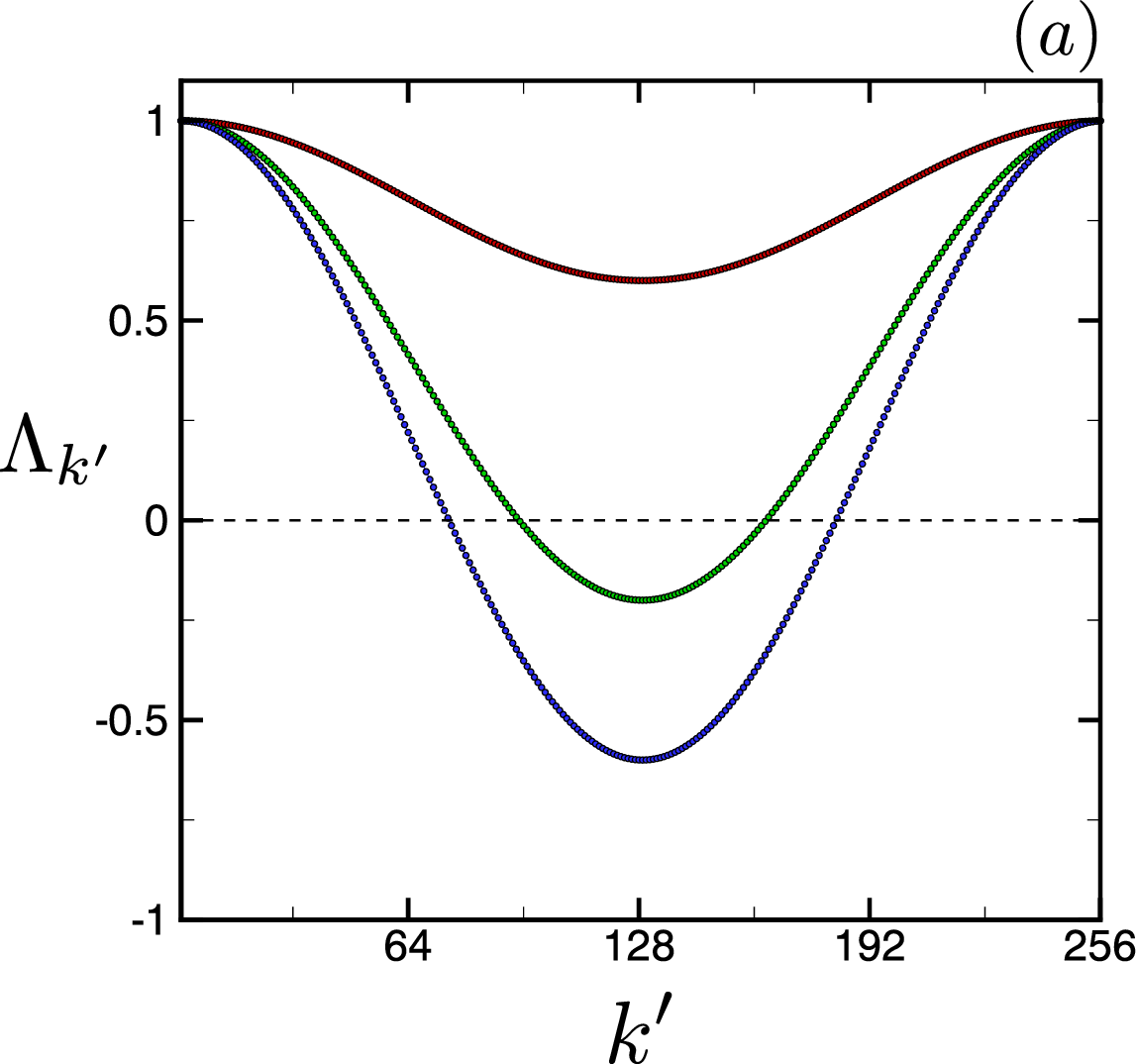}
\includegraphics[width=2.35in]{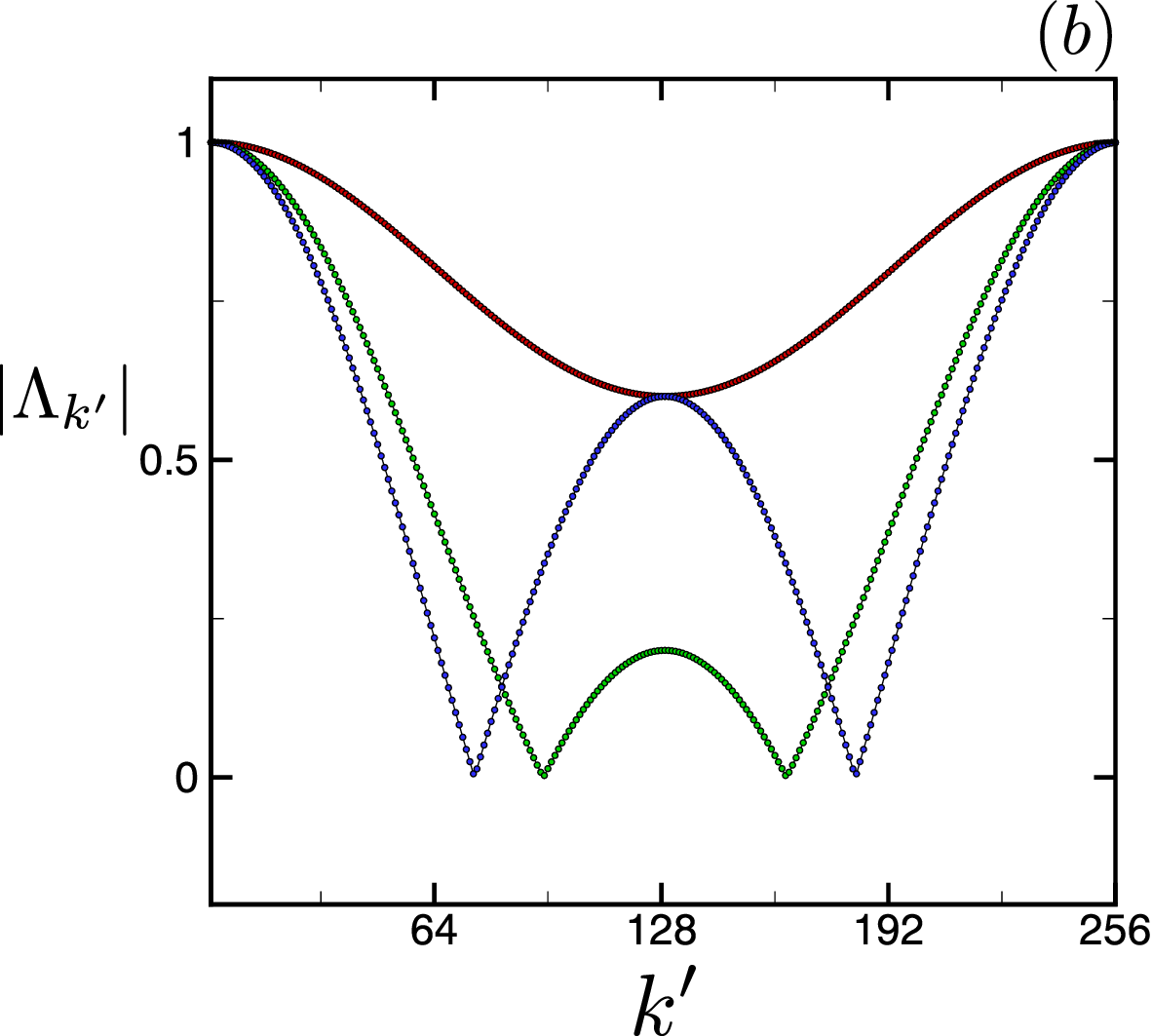}
\end{center}
\caption{The unsorted eigenvalues $\Lambda_{k'}$ of the diffusive coupling matrix $\mathbf{A}_c$ for $\epsilon\!=\!0.2$ (red), $\epsilon\!=\!0.6$ (green), and $\epsilon\!=\!0.8$ (blue). ($a$) The eigenvalues given by Eq.~(\ref{eq:eigenvalues-unsorted}). The minimum value crosses zero and becomes negative for  increasing values of $\epsilon$ where $\epsilon \!=\! 0.2$ (upper), $\epsilon \!=\! 0.6$ (middle), $\epsilon \!=\! 0.8$ (lower). ($b$) The variation of the absolute value of the eigenvalues. For $\Lambda_{k'} \!<\! 0$, which occurs for $\epsilon \!<\! 1/2$, two cusps emerge with a maximum occurring between the cusps. The magnitude of the maximum increases with increasing $\epsilon$.}
\label{fig:eigenvalues}
\end{figure}

The minimum value of the unsorted eigenvalues given by Eq.~(\ref{eq:eigenvalues-unsorted}) at $k' \!=\! N/2 \!+\! 1$ decreases linearly with increasing $\epsilon$ for $\epsilon \!\ge\! 0$. The minimum eigenvalue becomes zero at $\epsilon \!=\! 1/2$. For $\epsilon \!>\! 1/2$, the minimum becomes negative and there will be two zero crossings, with the location of the two zero crossings moving away in both directions from $k' \!=\! N/2 \!+\! 1$ with increasing $\epsilon$. This results in an increasing number of negative eigenvalues located at indices between the two zero crossings. Since $k$ is an integer there will not necessarily be an eigenvalue with a precisely vanishing value for a finite system for $\epsilon \!>\! 1/2$. The variation with diffusion strength of the eigenvalue index for the first eigenvalue whose magnitude is closest to zero, $k'_0(\epsilon)$, is given by 
\begin{equation}
k'_0(\epsilon) = \left\{ 1 + \frac{N}{2 \pi} \cos^{-1} \left( 1 - \frac{1}{\epsilon} \right) \right\}
\label{eq:k0}
\end{equation}
for $\epsilon \!\ge\! 1/2$. The $\{ \cdot \}$ notation is meant to indicate that it is necessary to round to the nearest integer to get the final result. The second minimum is at an index value located symmetrically about $N/2$ whose value is $N \!+\! 2 \!-\! k'_0(\epsilon)$ (these two eigenvalues will form a pair when sorted by magnitude). These trends are shown in Fig.~\ref{fig:eigenvalues}($a$) for several values of $\epsilon$.

However, it is the absolute value of the eigenvalues that is required in Eq.~(\ref{eq:lyap-theory}) to estimate Lyapunov spectrum. When considering the spectrum of $|\Lambda_{k'}|$,  the location of the eigenvalues with values closest to zero bound the region where $\Lambda_{k'}$ is negative for $\epsilon \!>\! 1/2$ which results in the emergence of two cusp structures. This is shown in Fig.~\ref{fig:eigenvalues}($b$).

The cusps occur at the indices where the magnitude of the eigenvalues are minimal and this location moves away from $k'\!=\! N/2 \!+\! 1$ with increasing values of $\epsilon$ for $\epsilon \ge 1/2$.  In the region between the cusps, centered at $k' \!=\! N/2 \!+\! 1$, the magnitude of the eigenvalues increase with increasing $\epsilon$ with a maximum value occurring at $k' \!=\! N/2 \!+\! 1$.  Two of these maxima are evident in Fig.~\ref{fig:eigenvalues}($b$) where the smaller maximum (green) is for $\epsilon \!=\! 0.6$ and the larger maxima (blue) is for $\epsilon \!=\! 0.8$. It is this increase in eigenvalue magnitude, between the cusps, that is ultimately responsible for the interesting behavior of the Lyapunov spectrum previously described for large negative Lyapunov exponents shown in Fig.~\ref{fig:lyap-spec-diff-with-epsilon}.

Now consider sorting the eigenvalues in order of descending magnitude to yield the spectrum $|\Lambda_k|$ for $k = 1, 2, \ldots, N$.   For $0 \! \le \! \epsilon \! < \! 1/2$, the tail of the Lyapunov spectrum decreases linearly. At $\epsilon \! = \! 1/2$ the presence of a vanishing eigenvalue results in the divergence of the largest negative Lyapunov exponent, $\lambda_N \rightarrow - \infty$, as indicated by Eq.~(\ref{eq:lyap-theory}).  This trend is evident in Fig.~\ref{fig:lyap-spec-diff-with-epsilon}($a$).

The growing region of increasing $\Lambda_{k'}$,  between the cusps of the unsorted eigenvalues, results in increasing values of the large negative Lyapunov exponents. This causes a kink in the sorted eigenvalue spectrum and therefore a kink in the predicted Lyapunov spectrum.

The kink in the predicted Lyapunov spectrum occurs at an index value of $k^*(\epsilon)$ where
\begin{equation}
k^*(\epsilon) = \left\{1 + \frac{N}{\pi} \cos^{-1} \left( 1 + \frac{2 (\epsilon - 1)}{\epsilon} \right) \right\}
\label{eq:kink}
\end{equation}
for $\epsilon \!>\! 1/2$ where, in this instance, $\{ \cdot \}$ indicates rounding down to the nearest integer. $k^*$ is the index of the Lyapunov exponent where the kink occurs, this is a single Lyapunov exponent which does not occur in a pair.  As a result of the kink, the magnitude of the Lyapunov exponents behind the kink, for $k \!>\! k^*$, \emph{increase} in magnitude for larger values of $\epsilon$ while still going to large negative values for $k \!=\! N$. This increase in the tail of the Lyapunov exponent spectrum is evident in our numerical results in Fig.~\ref{fig:lyap-spec-diff-with-epsilon}(b).

\begin{figure}[h!]
\begin{center}
\includegraphics[width=2in]{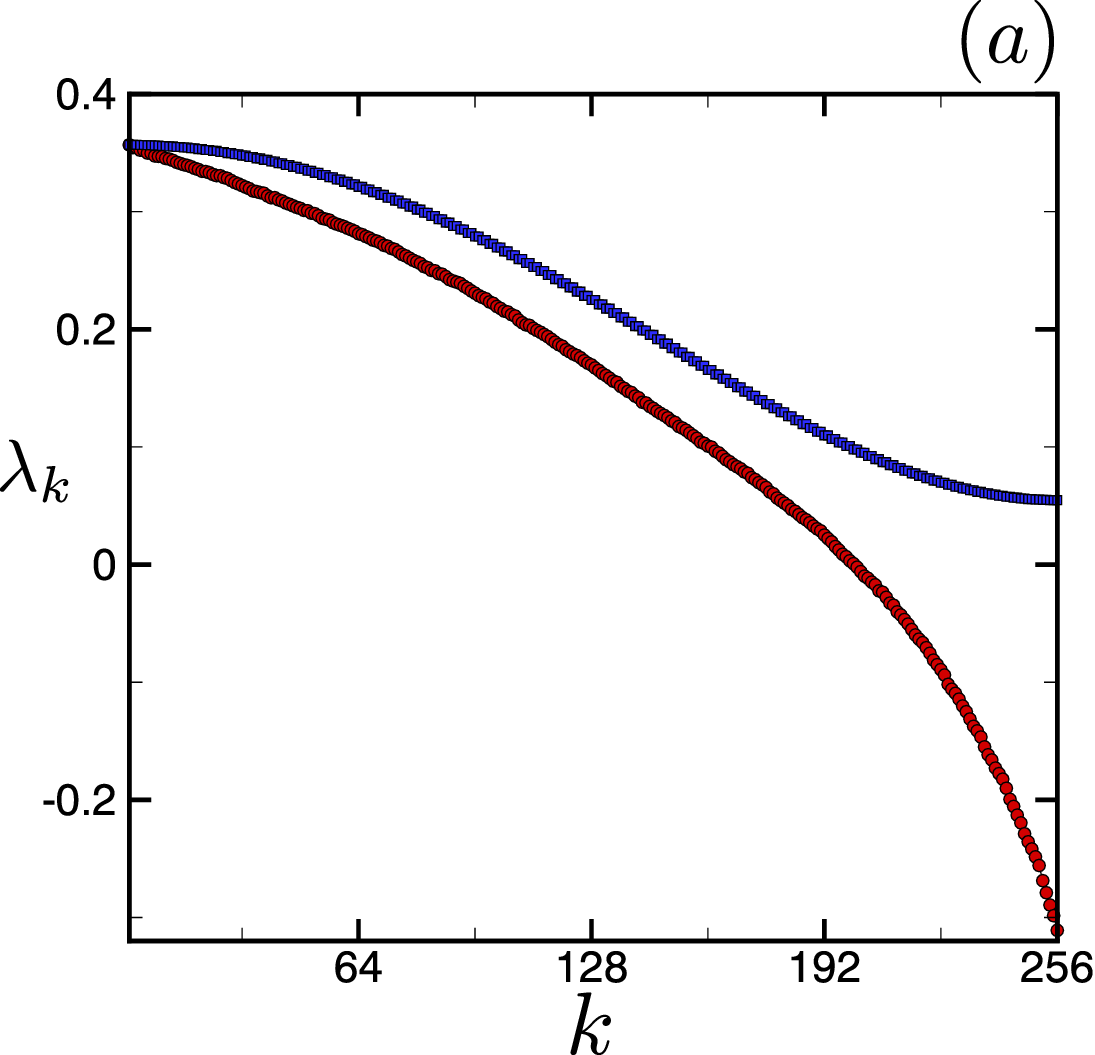}
\includegraphics[width=2.04in]{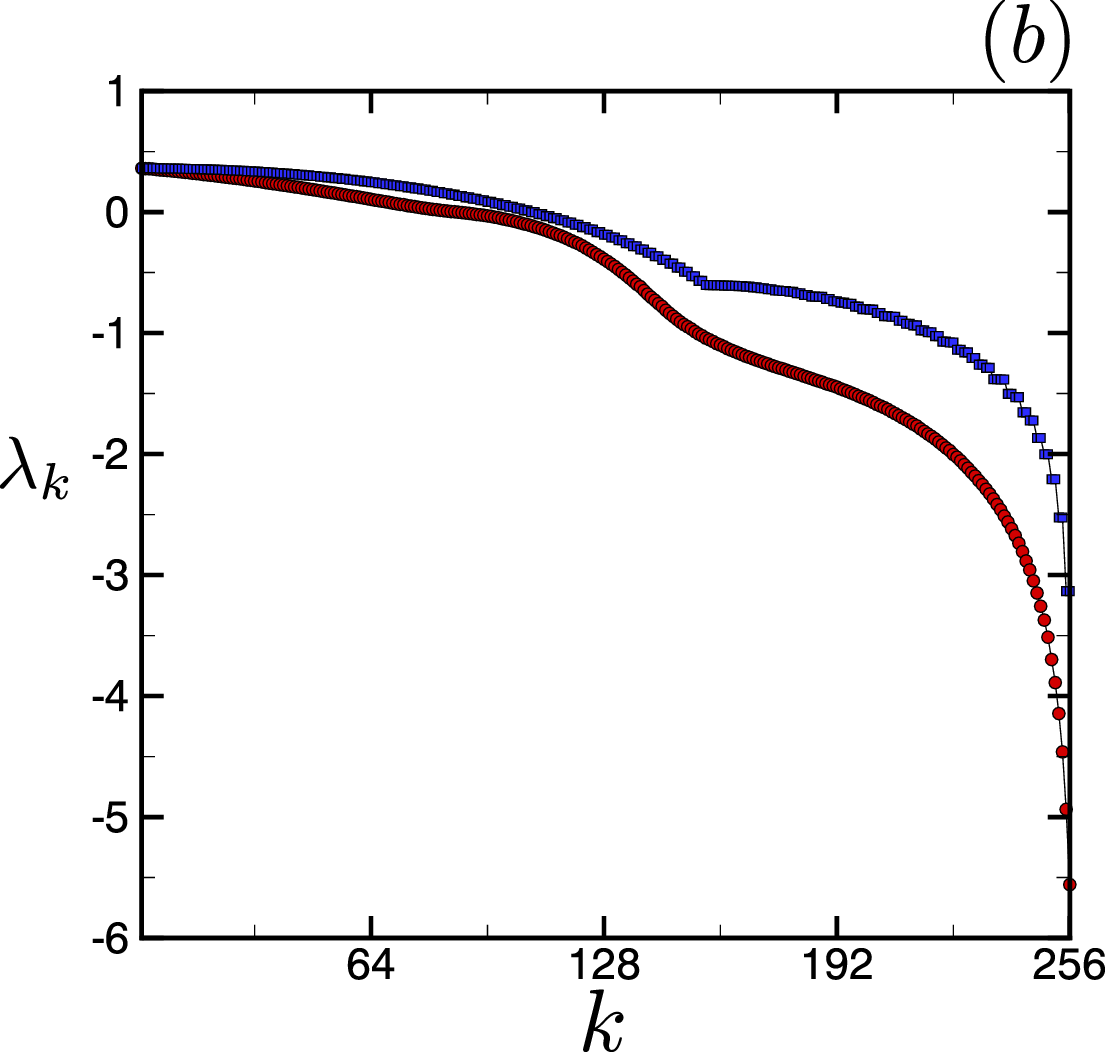}
\includegraphics[width=2.04in]{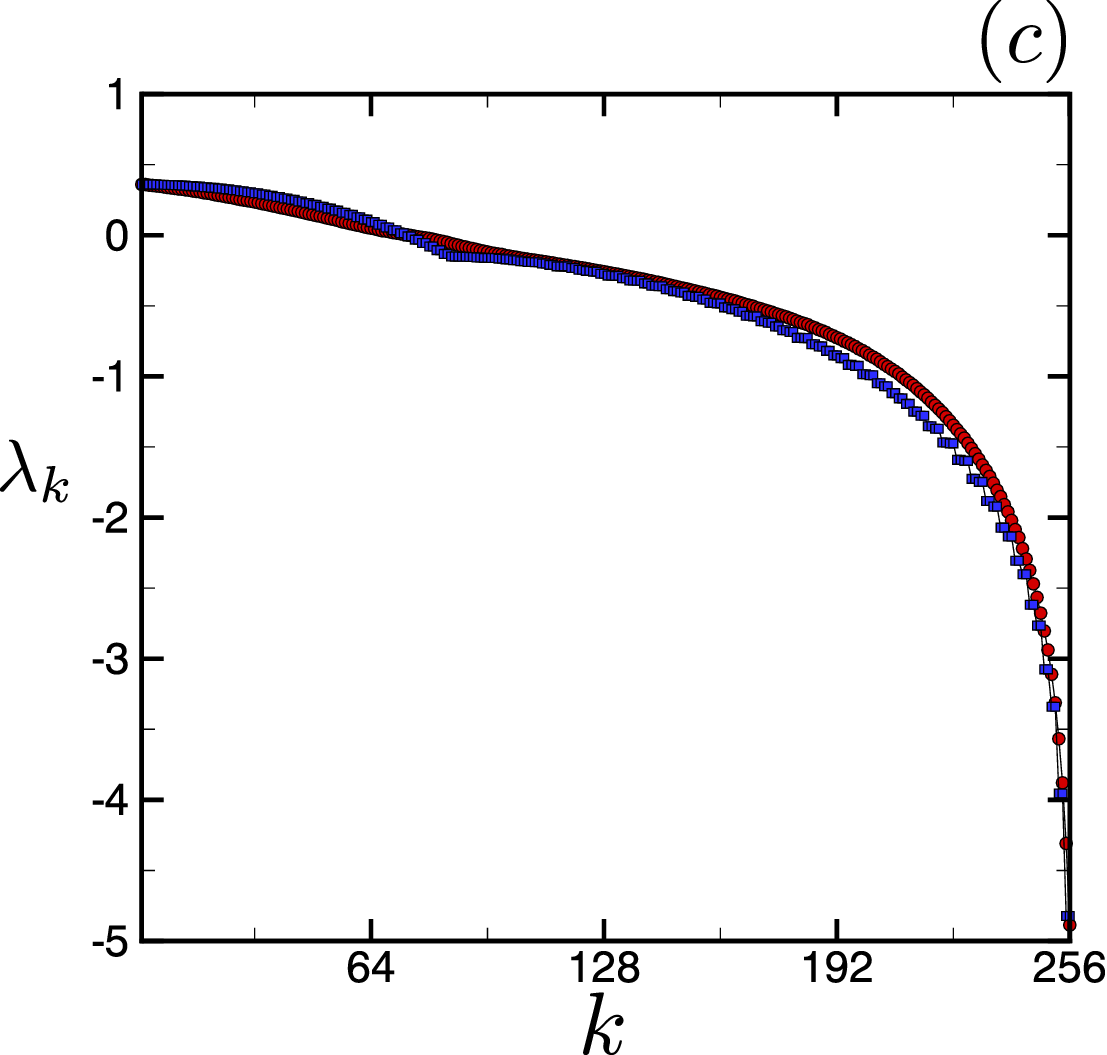}
\end{center}
\caption{Comparison of the Lyapunov spectra with the theoretical prediction of Eq.~(\ref{eq:lyap-theory}) for a range of diffusion strengths: ($a$)~$\epsilon \!=\! 0.2$, ($b$)~$\epsilon \!=\! 0.6$, ($c$)~$\epsilon \!=\! 0.8$ where theory (blue squares) and numerical simulation (red circles).}
\label{fig:lyap-comp}
\end{figure}

These ideas can be made more clear by a direct comparison of the theoretical estimation with the computed Lyapunov spectra from our simulations for several values of $\epsilon$.  The comparison of the Lyapunov spectra for these conditions is shown in Fig.~\ref{fig:lyap-comp}. The theoretical predictions determined by Eq.~(\ref{eq:lyap-theory}) are shown as the blue squares and the Lyapunov spectra are computed numerically from the lattice dynamics are shown by the red circles.  

For small diffusion strength where $\epsilon \!<\! 1/2$, as shown in Fig.~\ref{fig:lyap-comp}($a$) for $\epsilon \!=\! 0.2$, the theory over predicts the values of $\lambda_k$. In fact, the theory predicts that all of the Lyapunov exponents will be positive by underestimating the role of dissipation in the small dissipation limit. 

However, the theory does capture the rapid decrease of the Lyapunov exponents toward large negative values at large Lyapunov index $k$ with increasing values of $\epsilon$ as shown in Fig.~\ref{fig:lyap-comp}($b$)-($c$). The kink in the Lyapunov spectrum caused by the zero crossing the eigenvalues is now evident in the analytical results shown in Fig.~\ref{fig:lyap-comp}(b)-(c) by the blue, squares.  For Fig.~\ref{fig:lyap-comp}($b$), the kink occurs at an index value of $k^* \!=\! 156$, as given by Eq.~(\ref{eq:kink}). In Fig.~\ref{fig:lyap-comp}($c$) the agreement between theory and the numerics is very good where $k^* \!=\! 86$. However, as $\epsilon$ increases from $\epsilon \!=\! 0.8$ toward $\epsilon \!=\! 1$ the deviation again grows with the theoretical prediction capturing the general shape and trend of the spectra while also over predicting its magnitude. The usefulness of the estimate provided by Eq.~(\ref{eq:lyap-theory}) is not only in its quantitative comparison but in its ability to capture the overall trend and shape of the spectra.

The fractal dimension of the dynamics $D_\lambda$ can be calculated from the Lyapunov spectra using the Kaplan-Yorke formula
\begin{equation}
D_\lambda = j + \frac{\sum_{k=1}^j  \lambda_k} {| \lambda_{j+1}|}
\label{eq:fractal-dimension}
\end{equation}
where $j$ is the largest index such that the summation term is positive~\cite{kaplan:1979}.  In essence, $D_\lambda$ is the number of Lyapunov exponents that must be summed together to equal zero which corresponds to the ball of initial conditions in the tangent space that would neither grow nor shrink under the dynamics. As a result of the dependence on the Lyapunov exponents, the shape of the spectrum affects the value of $D_\lambda$ and the spatial coupling will affect the fractal dimension. 

\begin{figure}[h!]
\begin{center}
\includegraphics[width=3in]{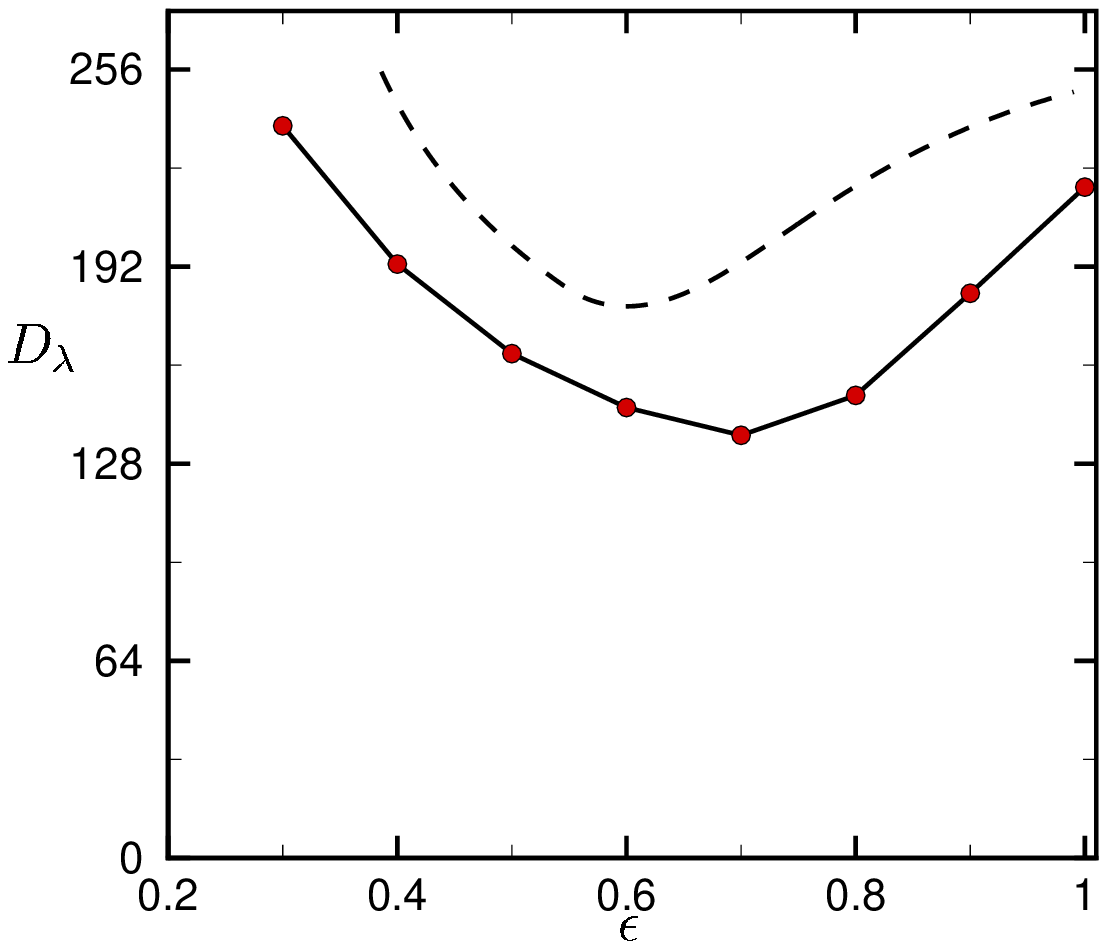}
\end{center}
\caption{Variation of the fractal dimension $D_\lambda$ with the strength of diffusion $\epsilon$. Circles are from numerical simulations. The dashed line is the theoretical prediction using the Lyapunov spectrum of Eq.~(\ref{eq:lyap-theory}) in Eq.~(\ref{eq:fractal-dimension}). $D_\lambda$ does not exist for low values of $\epsilon$ where the sum of the entire Lyapunov spectrum is positive.} 
\label{fig:dlamba-comp}
\end{figure}

The variation of $D_\lambda$ with $\epsilon$ is shown in Fig.~\ref{fig:dlamba-comp}.  The numerical results are shown by the red circles and the theoretical prediction is given by the dashed line.  For small values of $\epsilon$ the sum of the entire spectrum of $\lambda_k$ is positive and, in this case, the fractal dimension does not exist.  Overall, the increasing diffusion strength results in a decreasing value of $D_\lambda$ with a minimum value in our numerical simulations of $D_\lambda \!=\! 137.2$ at $\epsilon \!=\! 0.7$ which then increases to a final value of $D_\lambda = 217.8$ at $\epsilon \!=\! 1$. The theoretical prediction, shown by the dashed line, captures the general trend while over predicting the magnitude.

Lastly, we discuss the spatial structure of the eigenvectors $\vec{\xi}_k$ of the coupling matrix $\textbf{A}_c$ where  $k$ is the eigenvector index, $k \!=\! 1,2, \ldots, N$, and each eigenvector has $N$ components. We are assuming that the $\vec{\xi}_k$ have been ordered with the index $k$ that corresponds to the sorted eigenvalue spectrum with decreasing $\Lambda_k$ with increasing $k$.  In this ordering, the leading eigenvector $\vec{\xi}_1$ is the constant vector with a value of unity for each component $j$ where $j = 1,2, \ldots N$. For $k\!\ge\!2$ the eigenvectors are the Fourier modes $\vec{\xi}_{k} \!=\! \cos \left( \frac{\pi j k}{N} \right)$ and $\vec{\xi}_{k+1} \!=\! \sin \left( \frac{\pi j k}{N} \right)$ for $k = 2, 4, 6, \ldots, N$.

\subsection{Covariant Lyapunov vectors}
\label{section:clvs}

\begin{figure}[h!]
\begin{center}
\includegraphics[width=2.4in]{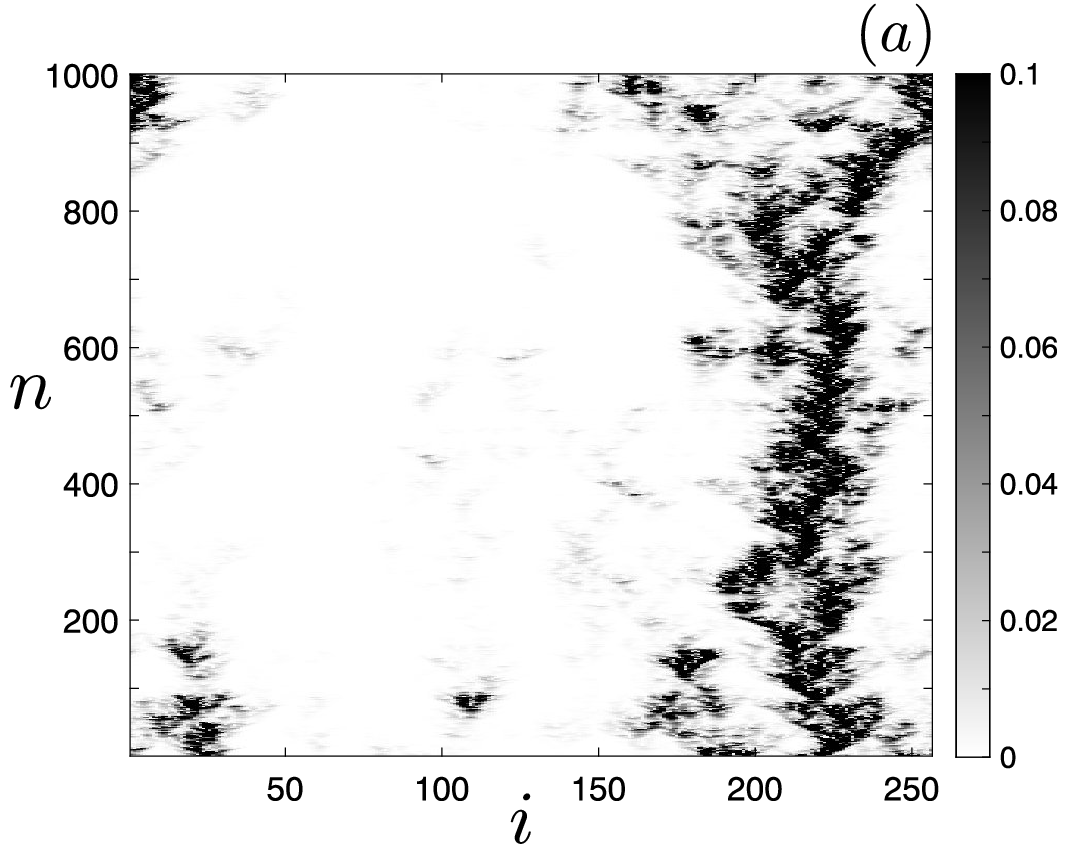} \hspace{0.4cm}
\includegraphics[width=2.4in]{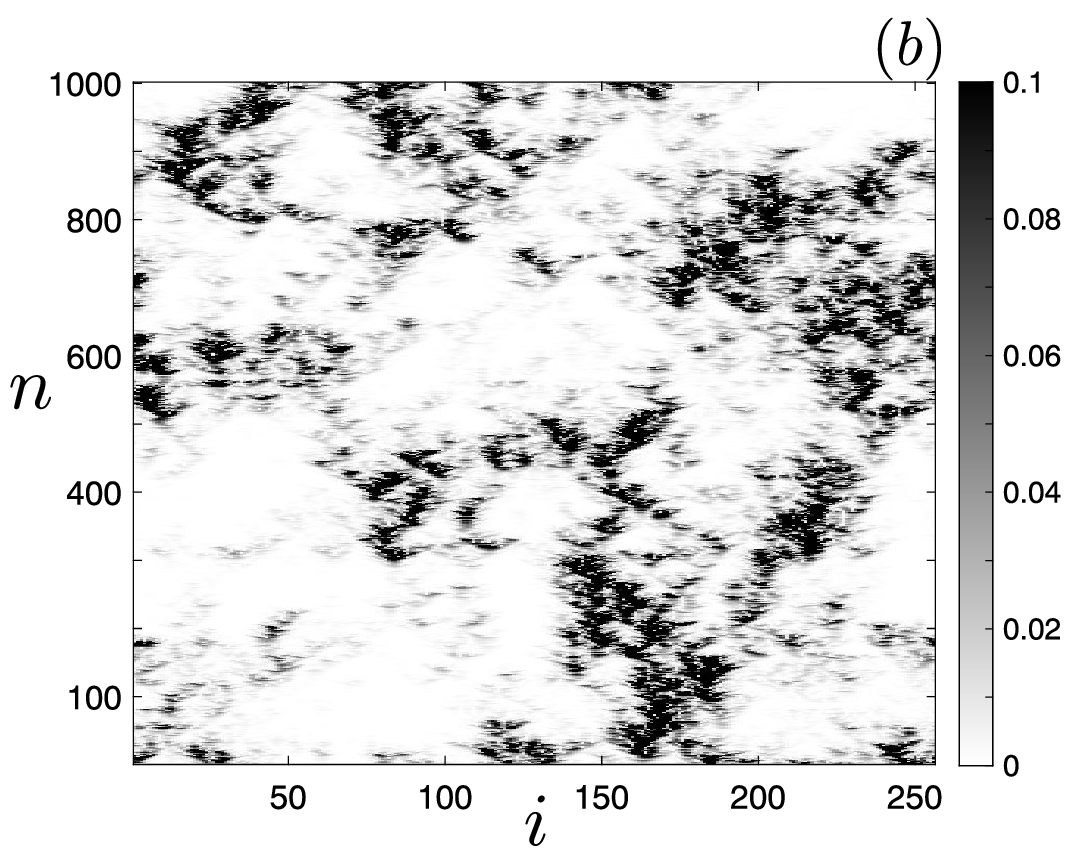} \\ \vspace{0.1cm}
\includegraphics[width=2.4in]{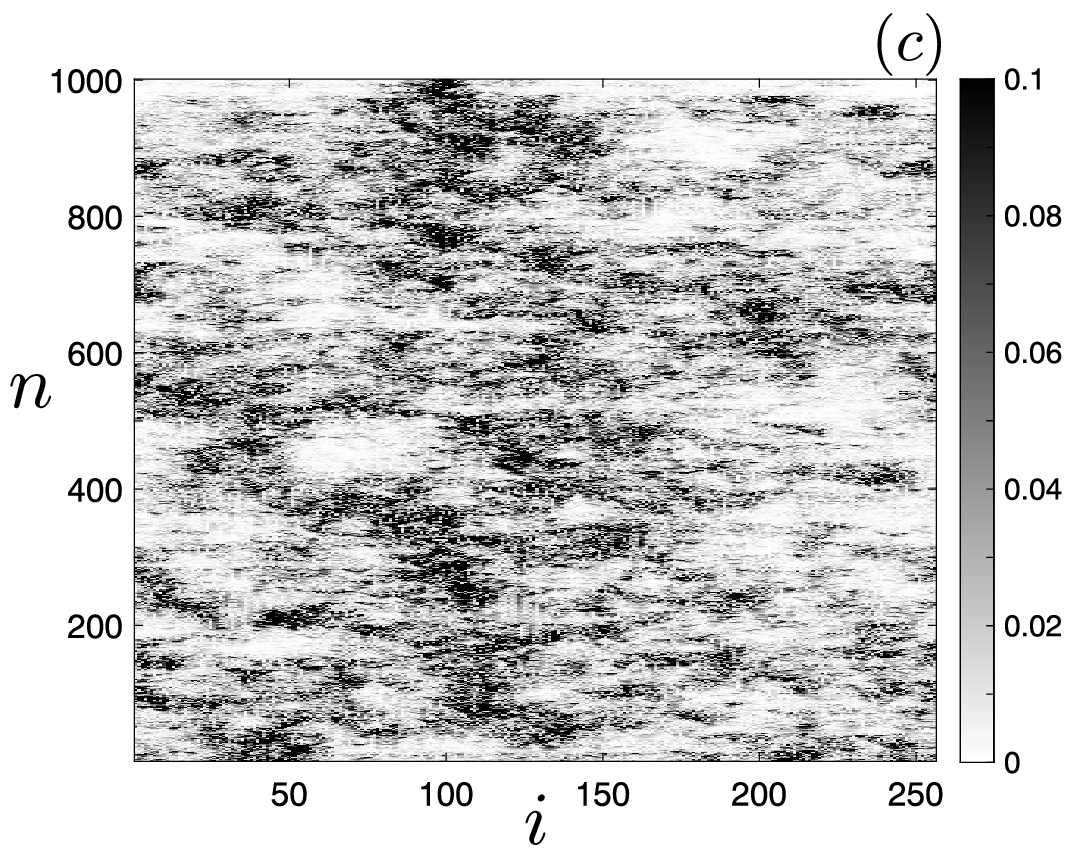} \hspace{0.4cm}
\includegraphics[width=2.4in]{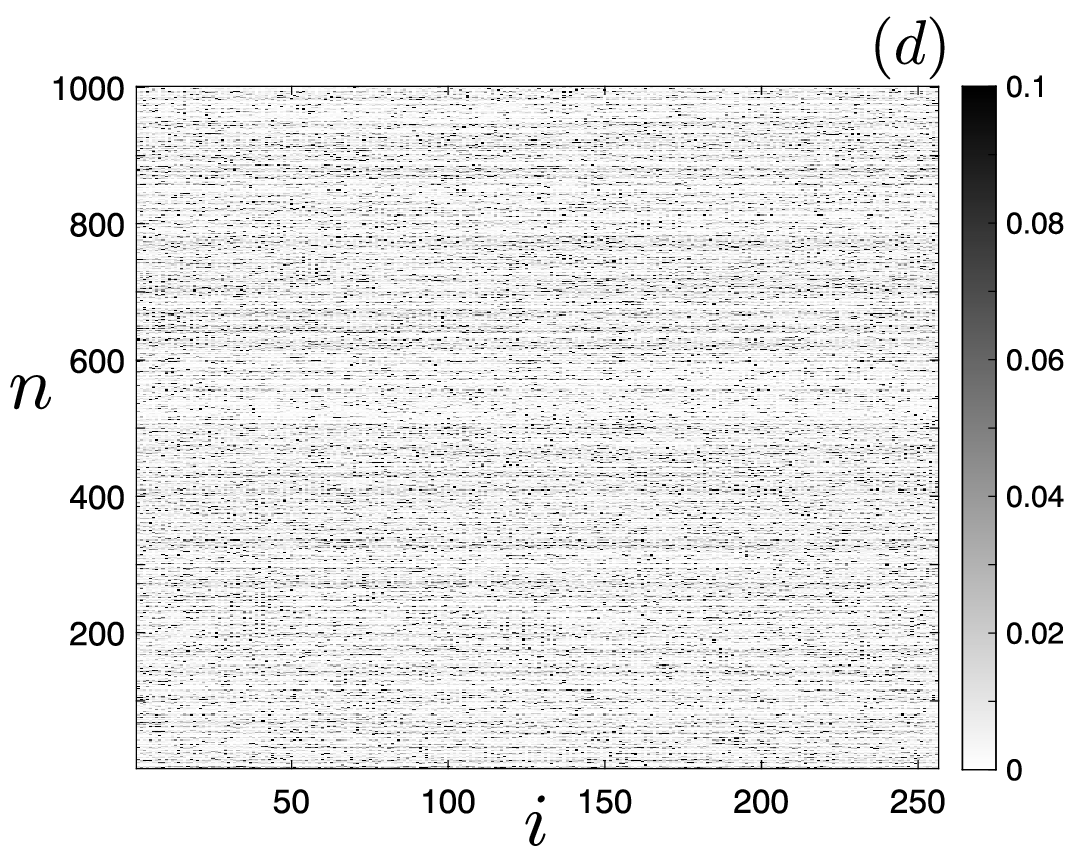}
\end{center}
\caption{Spacetime plots of the magnitude of the CLVs, $|\vec{v}^{(n)}_k|$, for $\epsilon \!=\! 0.8$ where $k$ is the Lyapunov index, $i$ is the component index for a particular CLV, and $n$ is the time step. ($a$) $k\!=\!1$ ($\lambda_1\!=\!0.360$), ($b$) $k\!=\!10$ ($\lambda_{10}\!=\!0.332$), ($c$) $k\!=\!100$ ($\lambda_{100}\!=\!-0.137$), ($d$) $k\!=\!256$ ($\lambda_{256}\!=\!-4.886)$.}
\label{fig:clvs-spacetime}
\end{figure}

Spacetime plots of the magnitude of several CLVs, $|\vec{v}_k^{(n)}|$, are shown in Fig.~\ref{fig:clvs-spacetime} for a diffusion strength of $\epsilon \!=\! 0.8$. Each panel shows results for a different CLV as given by its Lyapunov index $k$. The horizontal axis is the component $i$ of the $k$th CLV and the vertical axis is discrete time $n$.  The contours are of the  magnitude of the CLV with white representing small values and black representing large values as indicated by the color bar.

Figure~\ref{fig:clvs-spacetime}($a$) shows the spatiotemporal dynamics of the leading CLV where $k\!=\!1$. Localized structures of larger magnitude are evident.  Similar findings are shown in Figure~\ref{fig:clvs-spacetime}($b$) for $k\!=\!10$ which is also associated with a positive Lyapunov exponent. However, in this case the localized structure is more spatially spread out.  These trends continue for $k\!=\!100$ which is shown in Fig.~\ref{fig:clvs-spacetime}($c$) which corresponds with a negative Lyapunov exponent. The final CLV, $k\!=\!256$, is shown in Fig.~\ref{fig:clvs-spacetime}($d$) where the Lyapunov exponent has a large negative value and the localized structure is no longer present.

In order to quantify the spatial structure of the CLVs we have computed the inverse participation ratio $Y_2^{(k)}$ which is defined as
\begin{equation}
Y_2^{(k)} = \left< \sum_{i=1}^{N}  \left|v_{k,i}^{(n)} \right|^4 \right>
\end{equation}
where the angle brackets indicate an average over time~\cite{takeuchi:2011,mirlin:2000}. The summation is over the components $i$ of the $k$th CLV. A large value of $Y_2^{(k)}$ indicates significant localization and a small value indicates delocalization. The CLVs, $\vec{v}^{(n)}_{k}$, are normalized vectors with unit magnitude. The maximum possible value of $Y_2^{(k)}$ is unity corresponding to a CLV with a single component with a value of one and the remaining components all zero. The smallest possible value of $Y_2^{(k)}$ is $N^{-1}$ corresponding to a vector where every component has a value of $N^{-1/2}$. It is important to highlight that $Y_2^{(k)}$ does not capture the presence of spatial structure, but rather, the presence of significant localized values as quantified by large vector components of a normalized vector.

\begin{figure}[h!]
\begin{center}
\includegraphics[width=2.5in]{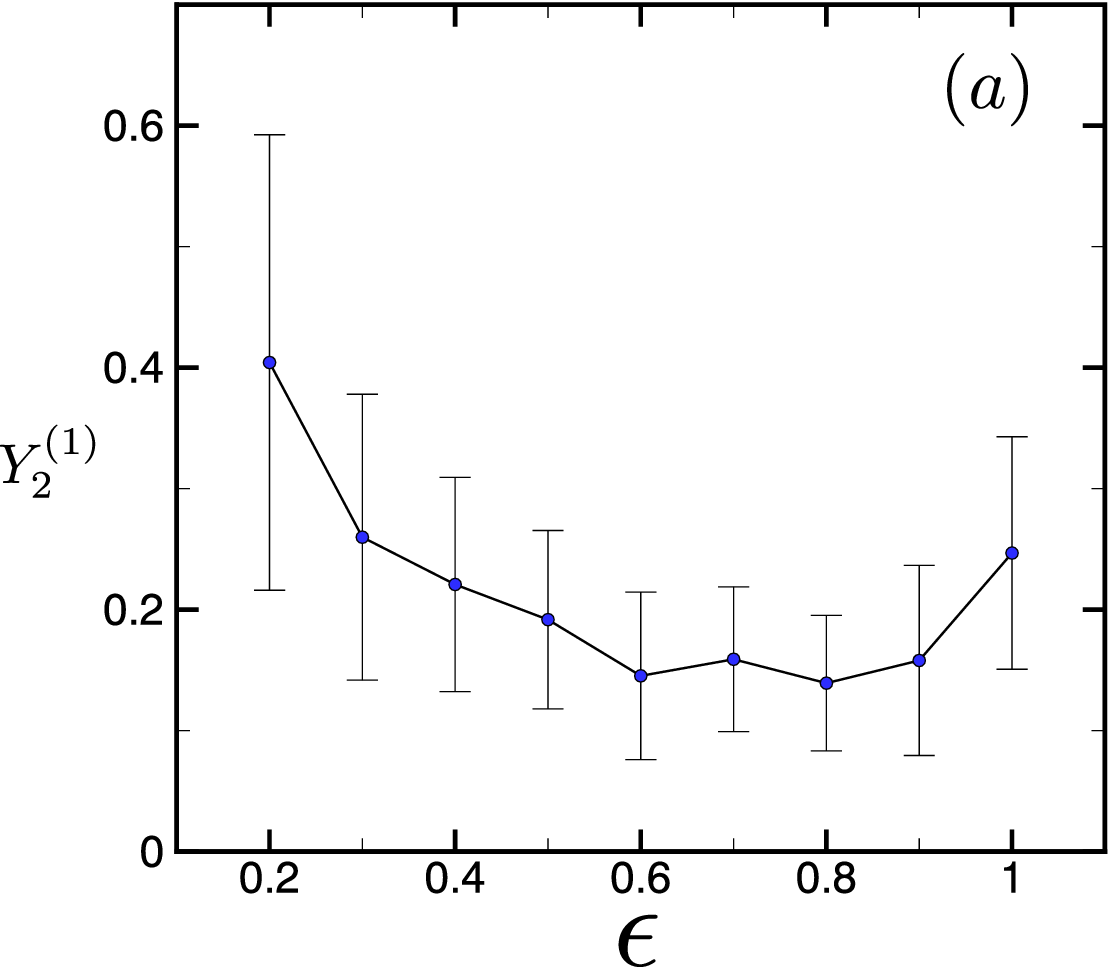} \hspace{0.4cm}
\includegraphics[width=2.55in]{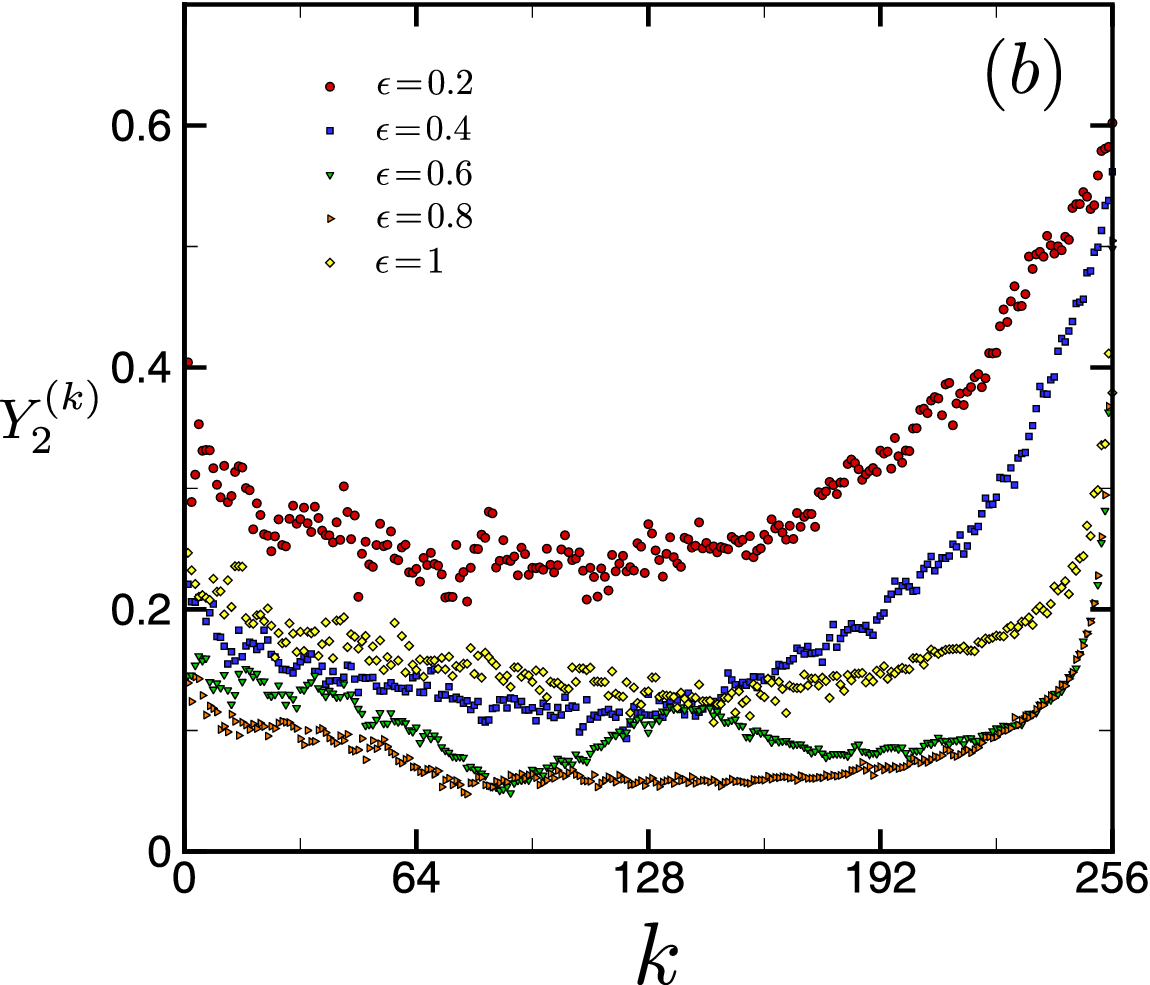}
\end{center}
\caption{The variation of the inverse participation ratio of the CLVs, $Y_2^{(k)}$, with the strength of diffusion $\epsilon$. ($a$)~The variation of the inverse participation ratio of the leading CLV ($k\!=\!1$). Data symbols are time averaged values and error bars are the standard deviation about the mean. ($b$)~Variation of $Y_2^{(k)}$ for the spectrum of CLVs for several values of $\epsilon$.}
\label{fig:ivpr-all}
\end{figure}

The variation of $Y_2^{(k)}$ is shown in Fig.~\ref{fig:ivpr-all} as a function of the diffusion strength. Figure~\ref{fig:ivpr-all}($a$) shows the variation of the localization for the leading CLV. The data symbols represent the time averaged values and the error bars are the standard deviation about the mean. The overall trend is a decrease in the localization of $\vec{v}_1$ for $0.2 \! \lesssim \! \epsilon \! \lesssim \! 0.8$ which is then followed by a slight increase.

Figure~\ref{fig:ivpr-all}($b$) shows the variation of the localization for the entire spectrum of CLVs over a range of diffusion strengths.  Over the entire range of diffusion strengths we have explored, the general trend is that the localization decreases gradually with $k$ until large $k$ at which point the localization increases rapidly to reach its maximum value for the $k=N_\lambda$. These trends are illustrated in Fig.~\ref{fig:clvs-spacetime} for the case of $\epsilon \!=\! 0.8$. The highly delocalized state is shown in Fig.~\ref{fig:clvs-spacetime}($d$). For the results shown in Fig.~\ref{fig:ivpr-all} all of the modes are physical modes.  The local maxima or bump feature in the localization results for the $\epsilon \!=\! 0.6$ case shown as green triangles in Fig.~\ref{fig:ivpr-all}($b$) can also be traced back the variation of the eigenvalues of the linear diffusion operator and the generation of the cusp feature shown in Fig.~\ref{fig:lyap-comp}($b$). 

\begin{figure}[h!]
\begin{center}
\includegraphics[width=3.2in]{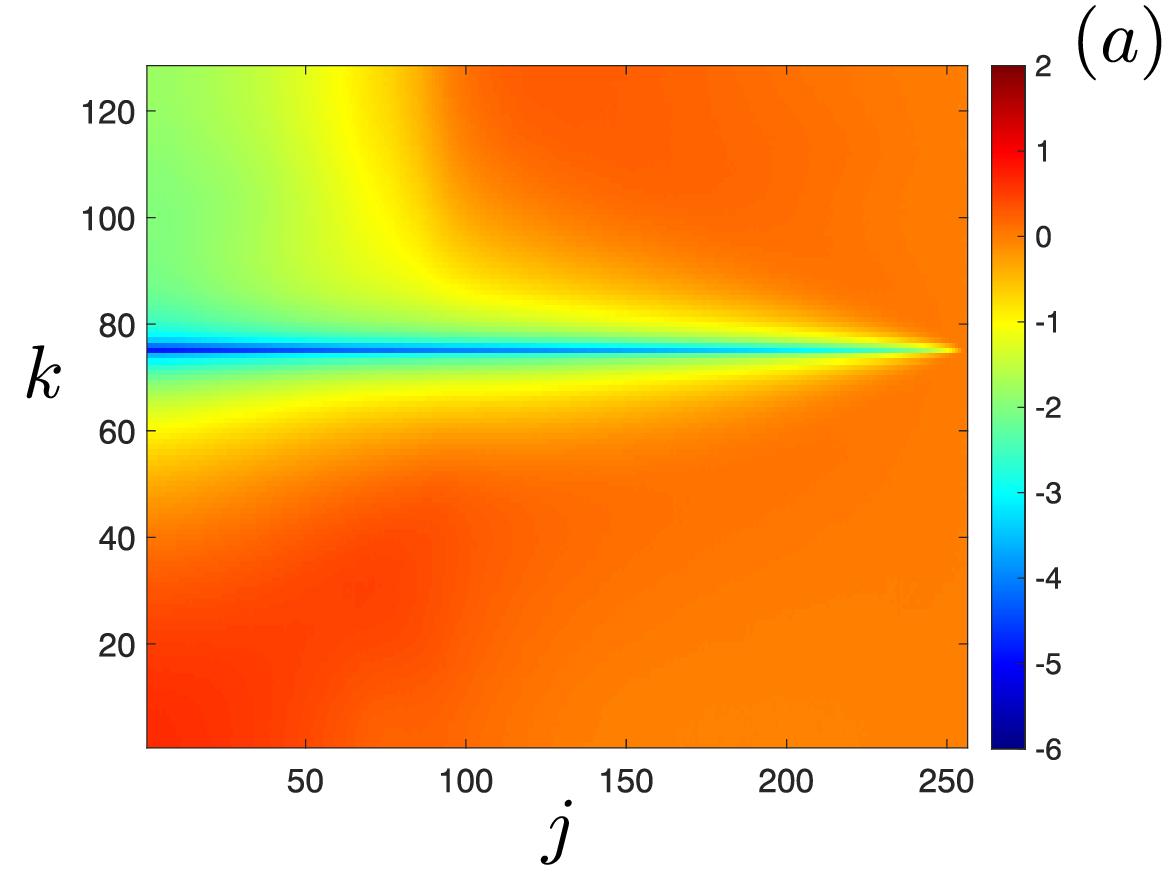} \hspace{0.4cm}
\includegraphics[width=3.0in]{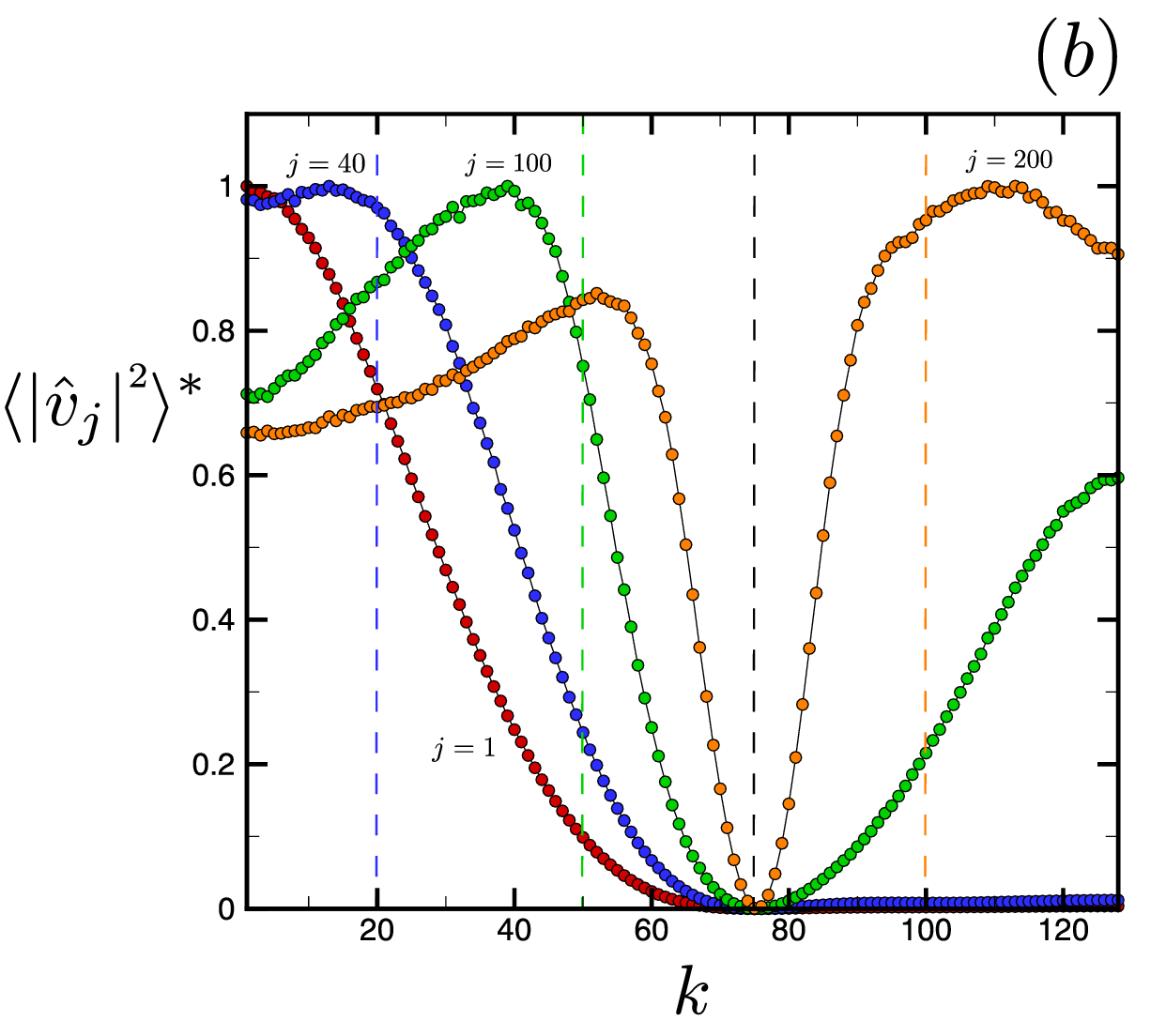}
\end{center}
\caption{Spatial variation of the CLVs and eigenvectors for $\epsilon \!=\! 0.8$. $(a)$ Spatial power spectrum of the CLVs, $\vec{v}_j$, where $j$ is the Lyapunov index and $k$ is the integer wavenumber.  Color contours indicate the time average of the magnitude of the discrete Fourier transform of the individual CLVs, $\langle |\hat{v}_j |^2 \rangle$, using a $\log_{10}$ scale where angle brackets indicate the time average. $(b)$ Detailed view of four spatial power spectra (labelled, color symbols) that are normalized to have a maximum value of unity, $\langle |\hat{v}_j |^2 \rangle^*$. Vertical dashed lines (color) indicate the wavenumber of the eigenvector corresponding to the CLV with data symbols of same color. For the $j$th CLV shown, the wavenumber of the corresponding eigenvector is $j/2$ when the eigenvectors are sorted by decreasing eigenvalue.  The black vertical line (dashed) indicates the index of the unsorted eigenvalue whose magnitude is closest to zero corresponding to the horizontal blue stripe in~$(a)$ occurring at $k_0 \!=\! 75$ (see blue curve of Fig.~\ref{fig:eigenvalues}(a)).}
\label{fig:clv-eig-spectra}
\end{figure}

Further insight into the spatial structure of the CLVs is provided by the spatial power spectrum. The spatial power spectrum is shown in Fig.~\ref{fig:clv-eig-spectra}(a) for $\epsilon \!=\! 0.8$ where $j$ is the Lyapunov index and $k$ is the integer wavenumber. The color contours are of the long-time average of the magnitude of the discrete Fourier transform $\langle |\hat{v}_j |^2 \rangle$. The hat notation indicates a discrete Fourier transform and the angle brackets indicate the time average. The color contours use a $\log_{10}$ scale and the time average is over a total time of $5 \times 10^4$ time units.  For the leading CLVs, $j \lesssim 50$, there is a significant magnitude at smaller wavenumbers, $k \lesssim 50$, which decreases to negligible values for $k \gtrsim 60$.  The horizontal blue stripe that occurs at $k_0=75$ corresponds with the index of the first unsorted eigenvalue whose magnitude is closest to zero. This can be seen as the first zero crossing of the blue curve in Fig.~\ref{fig:eigenvalues}(a) whose value of $k_0$ is given by Eq.~(\ref{eq:k0}).

The ability to predict the spectrum of Lyapunov exponents $\lambda_j$ using only the eigenvalues $\Lambda_j$ of the diffusive coupling operator, given by Eq.~(\ref{eq:lyap-theory}) and shown in Fig.~\ref{fig:lyap-comp}, suggests a connection between the CLVs, $\vec{v}_j$, and the eigenvectors $\vec{\xi_j}$. One indicator of this connection is the correspondence of the horizontal blue stripe of the spatial power spectrum of the CLVs shown in Fig.~\ref{fig:clv-eig-spectra}(a) with the index of the first zero eigenvalue of the diffusive coupling operator given by Eq.~(\ref{eq:k0}). Since the eigenvectors $\vec{\xi}_j$ are Fourier modes, their spectral magnitude is localized at the single wavenumber of the Fourier mode. For the case of $\epsilon \!=\! 0.8$ shown in Fig.~\ref{fig:clv-eig-spectra}, this indicates that eigenvector $\vec{\xi}_{k_0}$, with corresponding eigenvalue $\Lambda_{k_0} \approx 0$, would neither grow nor shrink by the action of the diffusive coupling operator. Figure~\ref{fig:clv-eig-spectra}(a) illustrates that, in fact, all of the CLVs do not grow or shrink at this value of $k_0$. This suggests that since the eigenvector at this wavenumber is unchanging the CLVs are not changing as well at this wavenumber.

We explore this further in Fig.~\ref{fig:clv-eig-spectra}(b) which shows a close-up view of the normalized spatial power spectra of four different CLVs as indicated by the labels. The spatial power spectra are normalized such that their maximum value is unity in order to focus upon their relative variations with the wavenumber $k$. For example, the green symbols show the variation of $\langle |\hat{v}_{100} |^2 \rangle^*$ with $k$ which corresponds to a vertical slice through Fig.~\ref{fig:clv-eig-spectra}(a) at $j\!=\!100$ and normalizing the maximum value to unity. For this case, the magnitude of the spatial power spectrum increases then reaches a minimum value at $k_0$ indicated by the vertical black dashed line (corresponding to the blue stripe in~(a)) and then increases for larger wavenumbers.  It is clear that the CLVs are not composed of only a few Fourier modes and that they all have a minimum value at $k_0$.

However, there is a strong contribution to the CLVs from the corresponding eigenvector. For example, the blue symbols of Fig.~\ref{fig:clv-eig-spectra}(b) show the variation of the spatial power spectrum for the 40th CLV.  The eigenvector pair with indices ($j$, $j+1$)  for even $j$ are given by $\vec{\xi}_j$ and $\vec{\xi}_{j+1}$. When the eigenvectors are sorted by decreasing value of the eigenvalues, each eigenvector of this pair has a wavenumber of $j/2$. Therefore, the corresponding eigenvector for the 40th CLV would have an integer wavenumber of 20. This is indicated by the vertical blue dashed line. It is evident that the spatial power spectrum of the CLV has a significant magnitude at this value of the wavenumber, suggesting the importance of this eigenvector. Similar results are shown by the green and orange vertical dashed lines for the 100th and 200th CLVs, respectively.

It has been shown for a number of important systems, including CMLs, ODEs, and PDEs, that it is possible to decompose the tangent space into physical and transient modes using the CLVs~\cite{yang:2009,kuptsov:2010,takeuchi:2011}. The physical modes are entangled, as indicated frequent near tangencies with one another in the tangent space, and are responsible for the dynamics of the system. In fact, the number of physical modes have been suggested as an estimate of the dimension of the inertial manifold~\cite{yang:2009,takeuchi:2011,ding:2016}.  The transient modes are hyperbolically isolated from all other modes, including other transient modes, and represent rapidly decaying, or extra, degrees of freedom. The decomposition of the tangent space can be quite complex, see for example Ref.~\cite{kuptsov:2010}, and its study remains an active area of research. 

One insightful way to explore this decomposition of the tangent space is by computing the violation of the domination of Oseledets splitting (DOS)~\cite{garnier:2006,pugh:2003,bochi:2005,yang:2008,takeuchi:2011}. The infinite-time Lyapunov exponents are guaranteed to be in descending order $\lambda_1 \! \ge \! \lambda_2 \! \ge \! \ldots$ However, over finite intervals of time, violations of this strict ordering may occur. It has been shown that such violations in the strict ordering, over finite times, occur when the CLVs are nearly tangent~\cite{pugh:2003,bochi:2005}. During these near tangencies, a small perturbation to the dynamics in the direction of the nearly tangent CLVs would affect both CLVs, this interaction is often described as the entanglement of the CLVs. In contrast, a perturbation to the dynamics in the direction of one of the hyperbolically isolated modes would affect only that rapidly decaying hyperbolically-isolated transient mode.

The violation of the DOS can be calculated by following the approach of Takeuchi \emph{et al.}~\cite{takeuchi:2011}. The $k$th finite time Lyapunov exponent $\tilde{\lambda}_k^\tau$ is defined as the value of the Lyapunov exponent determined over the interval of time from $t$ to $t\!+\!\tau$ where $\tau$ is a constant.  The difference between the finite time Lyapunov exponents with indices $k_1$ and $k_2$ can then be expressed as
\begin{equation}
\Delta \lambda_{k_1,k_2}^\tau = \tilde{\lambda}_{k_1}(t) - \tilde{\lambda}_{k_2}(t).
\end{equation}
For each time interval $\tau$ the presence or absence of a DOS violation is computed using $\Delta \lambda_{k_1,k_2}^\tau$. The violation of the DOS,  $\nu_{k_1,k_2}^{\tau}$, is then computed as the fraction of the time where $\Delta \lambda_{k_1,k_2}^\tau \!<\! 0$ which indicates that a violation has occurred. This can be written as
\begin{equation}
    \nu_{k_1,k_2}^\tau = \langle 1 - \mathcal{H} \left[ \Delta \lambda_{k_1,k_2}^\tau (t) \right] \rangle
    \label{eq:vdos}
\end{equation}
where $\mathcal{H}$ is the unit step function and the angle brackets indicate an average over time. For example, when a violation has occurred $\Delta \lambda_{k_1,k_2}^\tau \!<\! 0$ for that interval of time which yields a value of zero  from the evaluation of the step function to give a value of unity for the violation of the DOS. The final value of $\nu_{k_1,k_2}^\tau$ is simply the time average of these values computed for every time interval and for every possible pair of Lyapunov exponents.

For a physical mode, the violation of DOS should approach zero asymptotically with increasing values of $\tau$. However, for a transient mode there is a finite value of $\tau$ where the violations vanish. Since the variation of $\nu_{k_1,k_2}^\tau$ with respect to $\tau$ is not known analytically it can be difficult to determine the presence of transient modes.

\begin{figure}[h!]
\begin{center}
\includegraphics[width=3.1in]{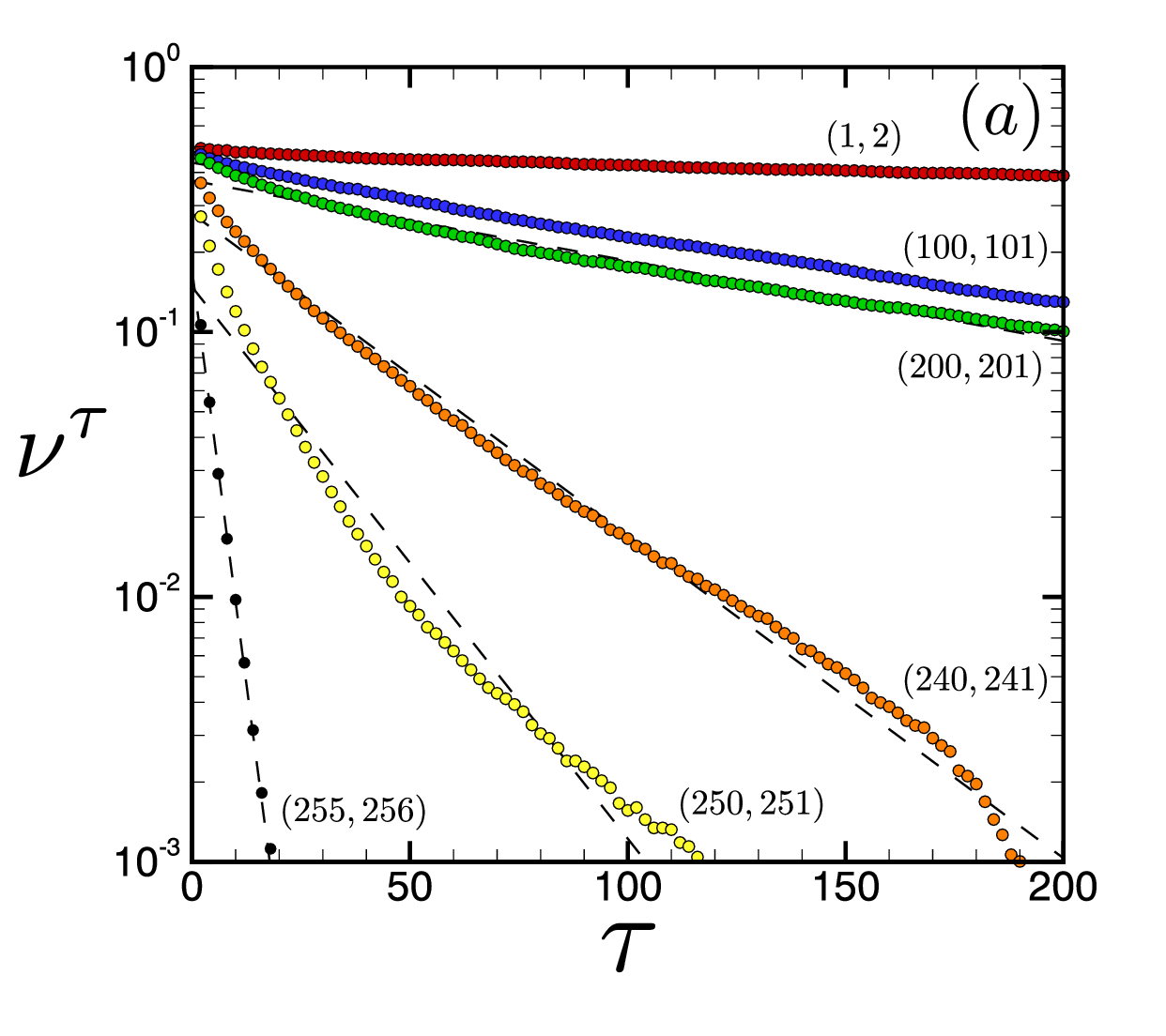}
\includegraphics[width=3.1in]{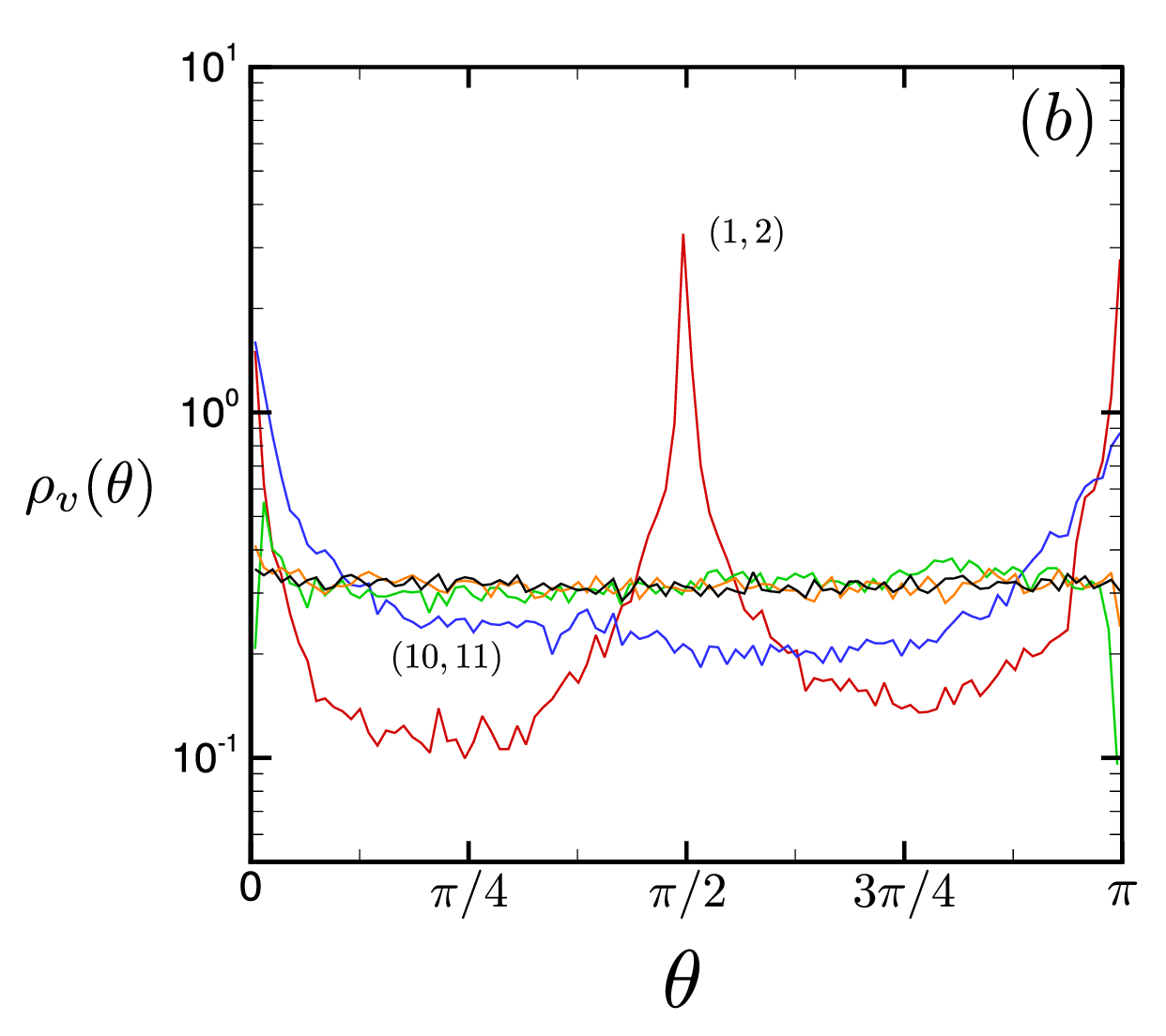}
\end{center}
\caption{The absence of transient modes for $\epsilon=0.7$ ($a$) The variation of the violation of DOS, $\nu^\tau$, with $\tau$ for several different representative pairs of CLVs indicated in parentheses.  The dashed lines are exponential curve fits through the data.  ($b$) The probability density function $\rho_v(\theta)$ of the angle $\theta$ between CLV pairs.  Pairs $(1,2)$ and $(10,11)$ are labelled and are given by the red and blue curves, respectively. Also shown are the unlabelled pairs $(100,101)$, $(250,251)$, $(255,256)$ in green, orange, and black respectively. For panels (a)-(b) the CLV's were calculated for $10^5$ time units.}
\label{fig:vdos-and-angle}
\end{figure}

As a representative example, in Fig.~\ref{fig:vdos-and-angle}(a) we show the variation of the violation of DOS with respect to $\tau$ for $\epsilon\!=\!0.7$ where we have computed the CLVs for $10^5$ time units. The violations are computed over a large range of $\tau$ where $2 \le \tau \le 200$ and results are shown using a semi-logarithmic scale for the six pairs of CLVs indicated by the labels. The black dashed lines represent exponential curve fits through the data. It is clear that the violations are significant over the entire range shown for first and second CLVs shown by the red symbols. For the vector pairs for increasing $k$, the decay in the violations is much faster with increasing $\tau$ yet they are all described well by exponential decay. From this variation alone the results suggest that $\nu_{k_1,k_2}^\tau$ can be described by exponential decay suggesting that these CLVs are physical modes. 

More insight can be gained by quantifying the statistics of the angle $\theta$ between pairs of CLVs. Figure~\ref{fig:vdos-and-angle}(a) shows the variation of the probability density function of the angle $\rho_v(\theta)$ with $\theta$ for several pairs of CLVs. The statistics of the angle between the first and second CLV are shown by the red curve where there is a peak near an angle of $\pi/2$ and also near an angle of zero (or $\pi$). The finite and significant probability of an angle $\theta$ near zero (or $\pi$) indicates the frequent occurrence of near tangencies between these vectors providing further support that these CLVs are physical modes. For the pair $(10,11)$, shown in blue, there is again a significant probability of near tangencies. The pairs (100,101), (250,251), and (255,256) are shown by the green, orange, and black curves respectively. All three pairs have a significant probability of near tangencies. As a result, all of the vectors shown are physical modes.

\begin{figure}[h!]
\begin{center}
\hspace{0.25cm} \includegraphics[width=2.15in]{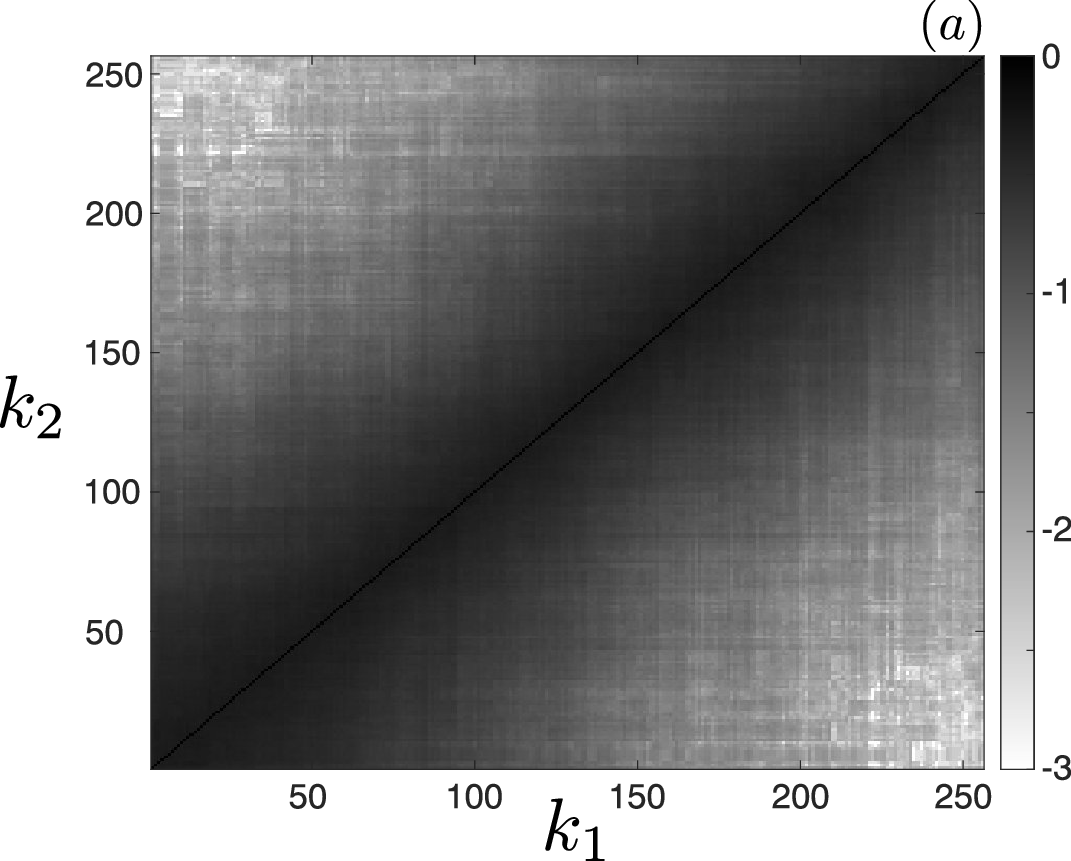}
\includegraphics[width=1.85in]{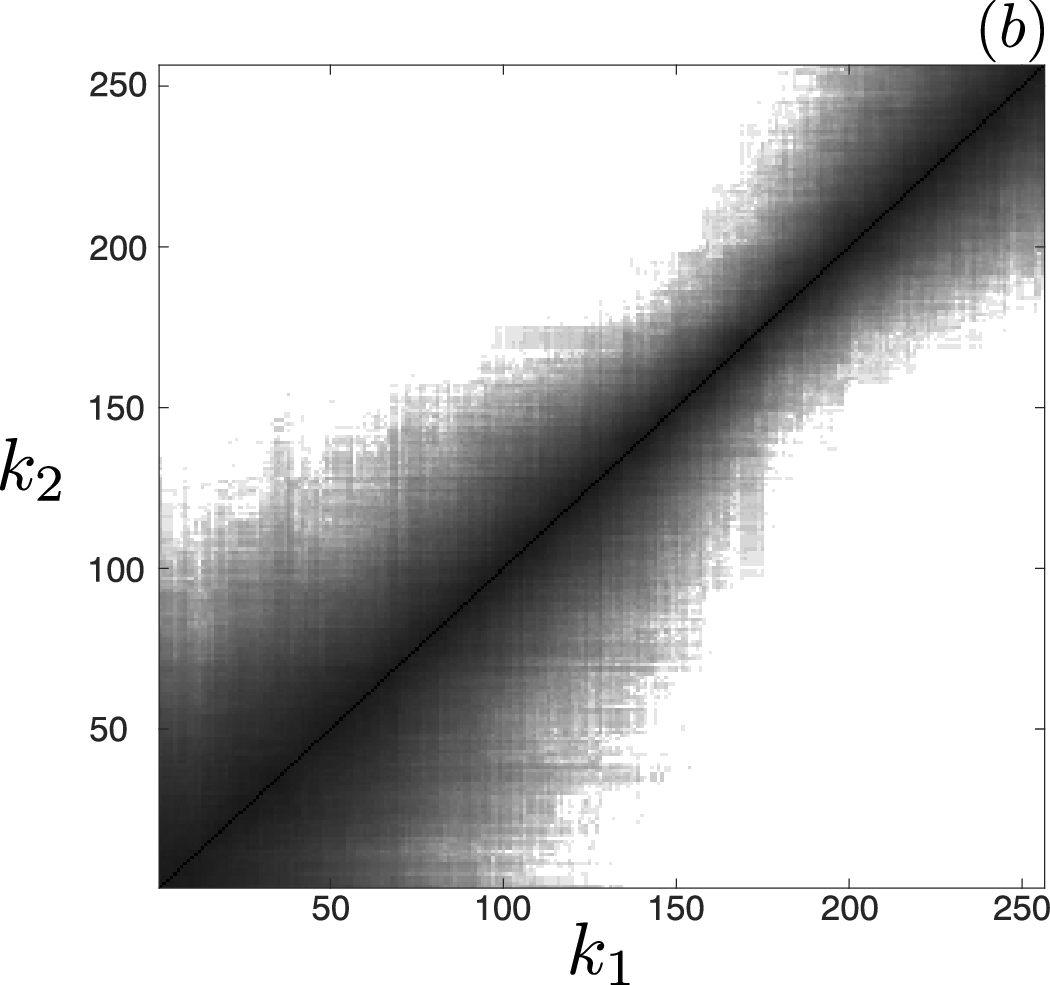}
\includegraphics[width=2.1in]{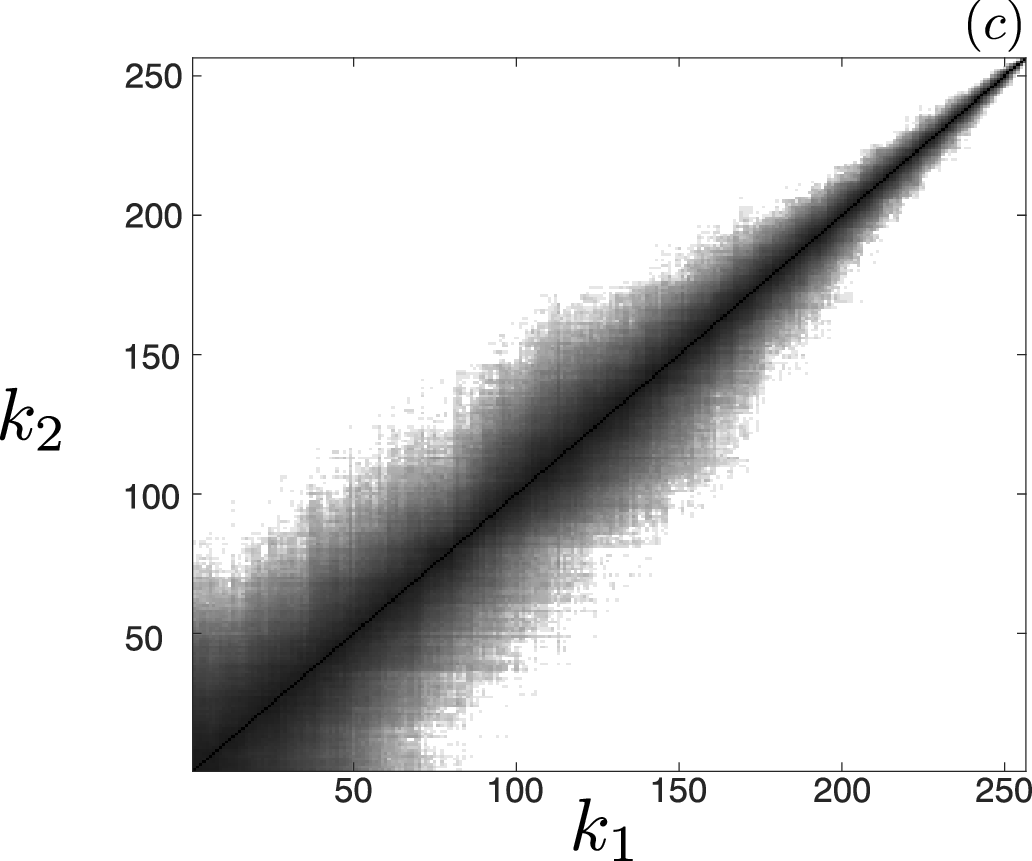}
\end{center}
\caption{Variation of the violation of the domination of Oseledets splitting, $\nu_{k_1,k_2}^\tau$, with the strength of diffusion $\epsilon$, see Eq.~(\ref{eq:vdos}). $k_1$ and $k_2$ are Lyapunov indices, the color contours are the $\log_{10}$ of the violation. Each panel uses the same color scale. ($a$)~$\epsilon \!=\! 0.2$, ($b$)~$\epsilon \!=\! 0.4$, ($c$)~$\epsilon \!=\! 0.8$. The amount of violation, or the amount of entanglement between the CLVs, reduces significantly with increasing $\epsilon$  over the range shown.}
\label{fig:vdos-with-epsilon}
\end{figure}

The variation of the violation of the DOS for all possible pairs of CLVs with the strength of diffusion is shown in Fig.~\ref{fig:vdos-with-epsilon} where we have used $\tau\!=\!5$. In light of the exponential decay of $\nu_{k_1,k_2}^\tau$ with $\tau$ for all CLV pairs shown in Fig.~\ref{fig:vdos-and-angle}(a), it is evident that the general structure of this image will not vary significantly with the choice of $\tau$.  The axes are the Lyapunov indices $k_1$ and $k_2$ of the two CLVs and the gray scale represents the magnitude of $\nu_{k_1,k_2}^\tau$ using a logarithmic scale where black indicates pure violations and white indicates the absence of violations. All panels use the color bar scale that is included in~($a$).

The three panels of Fig.~\ref{fig:vdos-with-epsilon} show the DOS violations for: ($a$) small diffusion strength, $\epsilon \!=\! 0.2$; ($b$) intermediate diffusion strength, $\epsilon\!=\!0.4$; and ($c$) for large diffusion strength $\epsilon\!=\!0.8$.  The main diagonal in each panel represents the special case where $k_1 \!= \! k_2$ which yields pure violation and therefore there is a black diagonal line running from the bottom left to the upper right. Additionally, the violation of the DOS plots are symmetric about the main diagonal since each computed value is a pairwise comparison.

Figure~\ref{fig:vdos-with-epsilon}(a) indicates that for small values of the diffusion strength there are significant violations. When only a small amount of diffusion is present the fluctuations in the values of the finite time Lyapunov exponents are large which yields significant deviations in their strict ordering. In the limit of no diffusion the lattice is simply $N$ uncoupled maps which would yield nearly pure violations for every pair of Lyapunov exponents.

As the diffusion increases the amount of violations of the DOS decreases significantly as shown in Fig.~\ref{fig:vdos-with-epsilon}($b$). In this case there are now a significant number of CLV pairs that do not show any violation as indicated by the regions of white space. This trend is continued for the large diffusion strength case that is shown in Fig.~\ref{fig:vdos-with-epsilon}($c$). As the diffusion strength increases the violations of the DOS are concentrated such that each CLV exhibits violations with respect to a smaller number of its nearest neighbors. The number of neighboring CLVs in which these violations occur can be determined as the vertical distance from the diagonal over which there are some violation indicated in grey. 

For all of the results shown in Fig.~\ref{fig:vdos-with-epsilon}, every CLV exhibits a violation with at least another CLV. In this case, the tangent space has not decomposed into physical and transient modes. As a result, every degree of freedom is contributing to the overall dynamics.

A close inspection of Fig.~\ref{fig:vdos-with-epsilon}(c) indicates that the amount of violations a CLV exhibits decreases with increasing Lyapunov index $k$. The leading CLVs, indicated by small $k_1$, have violations with a larger number of neighbors that the CLVs with large Lyapunov index. The general shape of the distribution of the violations is very similar to what has been found for Rayleigh-B\'enard convection~\cite{xu:2016} where spatial couplings due to diffusion and convection are significant.

As a measure of how the CLVs are entangled with their neighbors we have computed the degree of entanglement~\cite{barbish:2023} $\beta$.  The degree of entanglement is the fraction of the distinct pairs of CLVs, given by $k_1$ and $k_2$ where $k_1 \!\ne\! k_2$, which have a magnitude of the violation of $\nu_{k_1,k_2}^\tau \!\ge\! 0.001$. This corresponds to CLVs pairs which exhibit a violation of the DOS for 0.1\% of the time or more for the total duration of time we have explored.

\begin{figure}[h!]
\begin{center}
\includegraphics[width=3in]{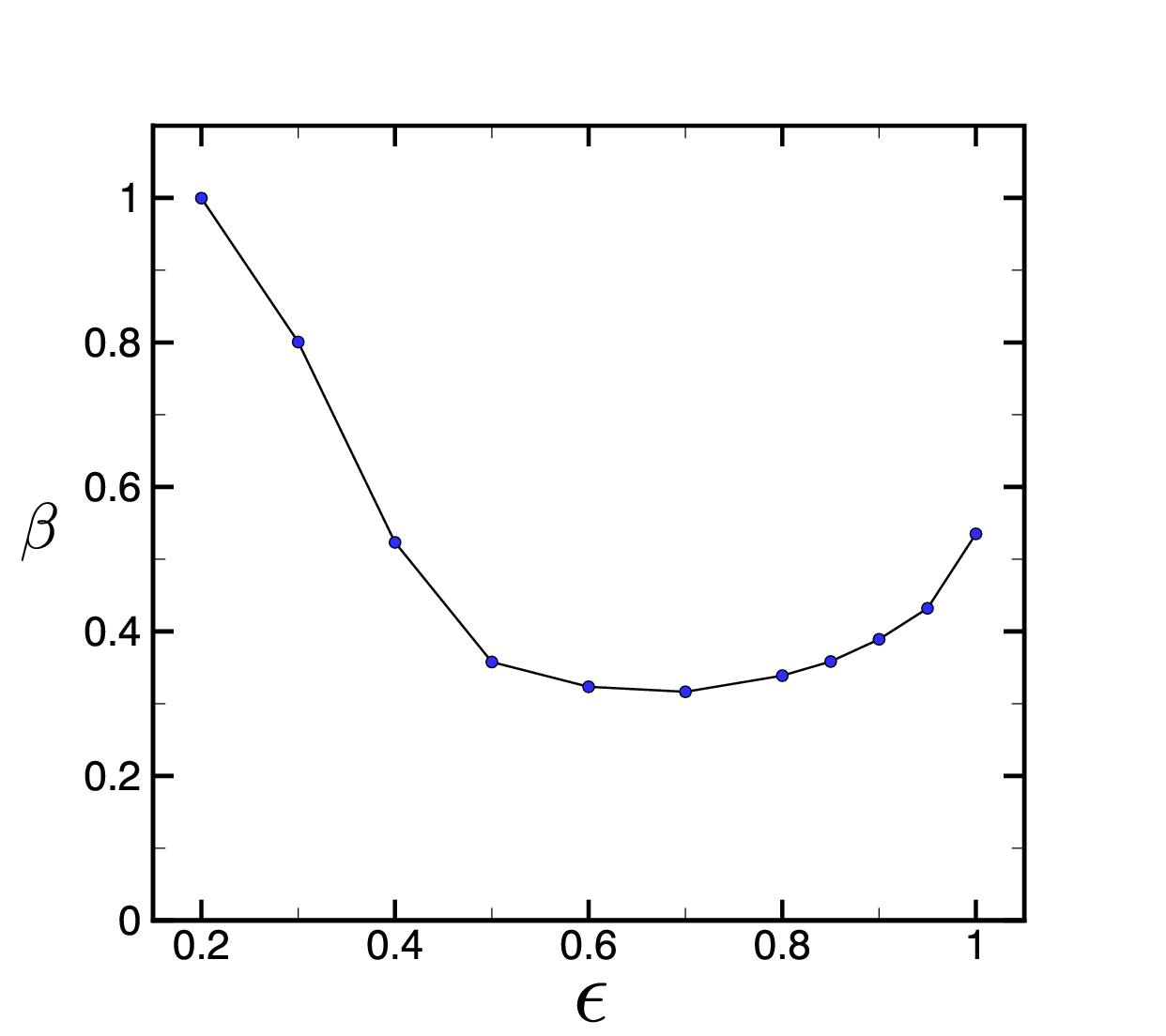}
\end{center}
\caption{The variation of the entanglement $\beta$ of the CLVs with diffusion strength $\epsilon$.}
\label{fig:entanglement}
\end{figure}

The variation of $\beta$ with $\epsilon$ is shown in Fig.~\ref{fig:entanglement}. The amount of entanglement decreases significantly until a moderate amount of diffusive coupling is present $\epsilon \!\gtrsim\! 0.5$. For larger value of diffusion $\beta$ slowly increases. This trend does not vary significantly with the choice of the threshold value of $0.1\%$

\section{Conclusion}
\label{section:conclusion}

We have investigated the role of diffusive coupling on the spatiotemporal chaos of a large one-dimensional lattice of quadratic maps. Our intention was to perform a fundamental study of the chaotic dynamics of a large system with locally generated chaos in the presence of a diffusive spatial coupling. The parameter space describing the dynamics of our chosen system of CMLs is vast and it is not our intention to provide an exhaustive description of the rich chaotic dynamics over this parameter space. Rather, we have carefully identified a range of parameters where the lattice dynamics is high dimensional, extensively chaotic, diffusively coupled, and computationally accessible. For these parameters we have conducted a broad study on the role of diffusion in spatiotemporal chaos where we have computed the full spectrum of CLVs. 

The presence of diffusive coupling has a significant effect upon the chaotic dynamics which we quantified in detail using the covariant Lyapunov vectors. The shape of the Lyapunov spectrum, and therefore the magnitude of the fractal dimension, is strongly affected by diffusive coupling. The variation of the Lyapunov spectrum and fractal dimension with the strength of the diffusive coupling can be described analytically using only the eigenvalues of the diffusive coupling operator and knowledge of the Lyapunov exponent of a single isolated map. It is important to highlight that the analytical description of the shape of the Lyapunov spectrum is independent of the particular map and only requires knowledge of the coupling operator. We also quantified the connection between the spatial features of the CLVs and the eigenvectors of the coupling operator.

Our investigation suggests that the diffusive coupling, for our lattices of quadratic maps, tends to organize the CLVs into a smaller bundle of entangled neighboring Lyapunov vectors where a neighboring Lyapunov vector is one whose Lyapunov index $k$ is close by. For the system we explored, this entangled bundle of CLVs occurs for all of the CLVs ($k = 1, 2, \ldots, N$) and we did not find a decomposition of the tangent space into physical and transient modes.

In fact, in the course of our investigation we performed an exploratory parameter sweep in terms of $r$ and $\epsilon$ and found the absence of the tangent space decomposition for all of the parameter values we tested.  This suggests that the lack of a decomposition is quite robust for this system and we speculate that this may be a result of the particular quadratic nonlinearity that we explored. However, it remains possible that the tangent space decomposition is present in some portion of the parameter space we did not probe. Of particular interest would be to explore the chaotic dynamics of the lattice for parameters where the fractal dimension is significantly smaller than the value of $D_\lambda \approx 128$ we have found in our study. Overall we do not have a firm understanding of the absence of the tangent space decomposition for the system we have explored and the general question of what is required by a dynamical system for the tangent space splitting to occur is an interesting direction to pursue further. This should be contrasted with the tangent space decomposition that has been found for diffusively coupled tent maps~\cite{takeuchi:2011,barbish:2023}.

A similar bundle of entangled CLVs has been found for chaotic Rayleigh-B\'enard convection in a periodic box domain~\cite{xu:2016}. However, in this case only 140 CLVs were computed due to computational expense and, as a result, it is not clear if perhaps a decomposition of the tangent space occurs at a higher Lyapunov index. The similarity between our CML results discussed here and the results from Rayleigh-B\'enard convection may also be a reflection of the significant diffusive transport mechanisms of momentum and heat in fluid convection and its effect upon the CLVs. Elucidating this possible connection further would be an interesting avenue of future study.

The approach we have used is quite general. It would be possible to explore different forms of nonlinearity, different types of spatial coupling, and higher dimensional lattices of maps to gain new insights into how these features affect the complex dynamics and to learn how these are reflected in the CLV description of the dynamics. The computational accessibility of lattices of maps remains a significant advantage allowing in depth and fundamental studies over a range of conditions using currently available computing and data storage resources which could otherwise quickly become prohibitive. 

\section*{Data Availability:}

The data that support the findings of this study are available from the corresponding author upon reasonable request.

\section*{Acknowledgments}

We acknowledge support from NSF fund number CMMI-2138055. Portions of the computations were performed using the Advanced Research Computing (ARC) center at Virginia Tech.

\end{document}